\documentclass[a4paper,fleqn]{cas-dc}
\hypersetup{
    colorlinks=true,
    citecolor=cyan,
    linkcolor=black,
    urlcolor=black
}
\usepackage{fancyhdr}
\fancypagestyle{plain}{% This will ensure changes even on chapter starting pages
  \fancyhf{} % Clear header and footer
}
\usepackage{hyperref}

\usepackage[inline]{enumitem}
\usepackage{amsmath,amsfonts}
\usepackage{graphicx} % For including images
\usepackage{apacite}
\usepackage{stfloats} % Add this in your preamble
\usepackage{color} % For textcolor
\usepackage{array}
\usepackage{pifont}
\usepackage{appendix}
\usepackage{textcomp}
\usepackage{colortbl}
\usepackage{xcolor}
\usepackage{booktabs}
\usepackage{url}
\usepackage{hhline}
\usepackage{multirow}
\usepackage{makecell}
\usepackage{float}
\usepackage{verbatim}
\usepackage{vcell}
\usepackage{color,soul}
\usepackage{longtable}
\usepackage{tabularray}
\usepackage{atbegshi}   %remove first blank page
\usepackage[linesnumbered,ruled,vlined]{algorithm2e}

\usepackage{array}
\usepackage{pifont}
\usepackage{textcomp}
\usepackage[dvipsnames]{xcolor}
\usepackage{url}
\usepackage{hhline}
\usepackage{multirow}
\usepackage{makecell}
\usepackage{mathtools}
\usepackage{float}
\usepackage{caption}
\usepackage{subcaption} % For subfigures and subcaptions
\usepackage{adjustbox}
\usepackage{colortbl}
\usepackage{verbatim}
\usepackage{vcell}
\usepackage{tikz}
\usepackage{cite}
\usepackage{orcidlink}
\usepackage{longtable}
\usepackage{algorithm}
\usepackage{comment}
\usepackage{verbatim}

\AtBeginShipoutNext{\AtBeginShipoutDiscard}   %remove first blank page
\makeatletter
\def\ps@plain{%
  \let\@mkboth\@gobbletwo
  \def\@oddhead{\hfil\thepage} % Page number at the top right for odd pages
  \def\@evenhead{\thepage\hfil} % Page number at the top right for even pages
  \let\@oddfoot\@empty % Empty footer
  \let\@evenfoot\@empty % Empty footer
}
\makeatother
\usepackage{natbib}
\usepackage{balance}

\def\tsc#1{\csdef{#1}{\textsc{\lowercase{#1}}\xspace}}
\tsc{WGM}
\tsc{QE}
\makeatletter
\renewcommand\subsection{\@startsection{subsection}{2}{\z@}%
   {-3.25ex\@plus -1ex \@minus -.2ex}%
   {1.5ex \@plus .2ex}%
   {\normalfont\large}} % Adjust font size and family here
\renewcommand\subsubsection{\@startsection{subsubsection}{3}{\z@}%
   {-3.25ex\@plus -1ex \@minus -.2ex}%
   {1.5ex \@plus .2ex}%
   {\normalfont\normalsize}} % Adjust font size and family here
\makeatother

\begin{document}

\let\WriteBookmarks\relax
\def\floatpagepagefraction{1}
\def\textpagefraction{.001}
\renewcommand{\figurename}{Fig.}\

\begin{comment}
% Title page
\begin{titlepage}
    \centering
    \vspace*{5cm}
    {\huge\bfseries PQBFL: Post-Quantum and Blockchain-based
Protocol for Federated Learning\par}
    \vspace{1.5cm}
    {\Large\textbf{Hadi Gharavi \quad Jorge Granjal \quad Edmundo Monteiro }\par}
    \vspace{1cm}
    {\large CISUC, Department of Informatics Engineering, University of Coimbra, Coimbra, Portugal\par}
    \vspace{0.5cm}
    {\large \{hgharavi, jgranjal, edmundo\}@dei.uc.pt\par}
    \vfill
\end{titlepage}
\end{comment}

% Start the main document on the second page

\setcounter{page}{1}
\title [mode = title]{PQBFL: A Post-Quantum Blockchain-based
Protocol for Federated Learning}  
\ead{E-mail addresses: hgharavi@dei.uc.pt (H. Gharavi), jgranjal@dei.uc.pt (J. Granjal),
edmundo@dei.uc.pt (E. Monteiro).}
\author{\textcolor{black}{Hadi Gharavi\textsuperscript{\hyperlink{corrauth}{*}}, Jorge Granjal, Edmundo Monteiro}}
\affiliation[]{organization={University of Coimbra, Centre for Informatics and Systems of the University of Coimbra, Department of Informatics Engineering}, country={Portugal}}

\cortext[1]{Corresponding author at CISUC, Department of Informatics Engineering, University of Coimbra, Portugal}

\begin{abstract}
One of the goals of Federated Learning (FL) is to collaboratively train a global model using local models from remote participants. 
However, the FL process is susceptible to various security challenges, including interception and tampering models, information leakage through shared gradients, and privacy breaches that expose participant identities or data, particularly in sensitive domains such as medical environments. 
Furthermore, the advent of quantum computing poses a critical threat to existing cryptographic protocols through the Shor and Grover algorithms, causing security concerns in the communication of FL systems.
To address these challenges, we propose a Post-Quantum Blockchain-based protocol for Federated Learning (PQBFL) that utilizes post-quantum cryptographic (PQC) algorithms and blockchain to enhance model security and participant identity privacy in FL systems.
It employs a hybrid communication strategy that combines off-chain and on-chain channels to optimize cost efficiency, improve security, and preserve participant privacy while ensuring accountability for reputation-based authentication in FL systems.
The PQBFL specifically addresses the security requirement for the iterative nature of FL, which is a less notable point in the literature. 
Hence, it leverages ratcheting mechanisms to provide forward secrecy and post-compromise security during all the rounds of the learning process.  
In conclusion, PQBFL provides a secure and resilient solution for federated learning that is well-suited to the quantum computing era.
\end{abstract}


%\item 
%\item 
%\item 

% Keywords

\maketitle

\setcounter{page}{1}

\section{Introduction}
\label{Introduction}

The emergence of quantum computing in the near future poses a significant threat to conventional cryptographic standards, putting secure communication and sensitive data at risk. 
Shor \citep{shor1994algorithms} and Grover \citep{grover1996fast} are quantum-based algorithms that can easily compromise current cryptography algorithms. 
Shor's algorithm efficiently finds prime factors of large numbers, breaking RSA encryption, whereas Grover's algorithm speeds up brute-force attacks on symmetric-key cryptography, reducing their security. 
Although quantum computers are currently not powerful enough to be considered a serious threat, the real concern is the ``Harvest Now, Decrypt Later" (HNDL) attack, in which adversaries collect encrypted data with the intention of decrypting it once quantum computing becomes sufficiently powerful.  
Therefore, it is necessary to develop post-quantum cryptographic (PQC) protocols for various applications that can withstand quantum-fueled attacks \citep{Post-Quantum_Blockchain_Security_for}.  

In the context of Federated Learning (FL), in which multiple participants collaboratively train a global model on an aggregator server by sending their local models, the confidentiality, integrity, and authenticity of the model data are crucial to the duration of the learning process. 
It is also possible that the model data value does not decrease over time, making the HNDL attack reasonable for FL systems.  
Furthermore, there are additional concerns about privacy, availability, and threats related to machine learning, such as model or data poisoning attacks.  
These attacks affect the accuracy of the final aggregated model by injecting false data or manipulating local models. 
Therefore, federated learning schemes use blockchains as security, transparency, distribution, and incentive aid \citep{Towards_blockchain-based,Securing_federated}.  
The blockchain can serve as an immutable record for all contributions in the training model, enabling the easy identification and prevention of malicious activities that increase the resilience of FL systems to such attacks. 
Blockchain can also contribute to the collective training process without revealing a participant's identity. 
This allows participants to engage in the process through pseudonyms, whereas contributions can be traced back to accountability.  
 
The future-looking security and user privacy-preserving requirements in FL motivated us to present a Post-Quantum Blockchain-based protocol for Federated Learning (PQBFL). 
The PQBFL protocol considers the critical needs of an FL framework to offer a post-quantum security solution with advancements in quantum computing.  Moreover, leveraging blockchain characteristics provides mechanisms for decentralization, privacy preservation, and reputation management of FL projects.  Therefore, our main contributions to the PQBFL are summarized as follows.  
 \begin{enumerate}     
    \item The PQBFL provides a hybrid solution using conventional and post-quantum cryptographic primitives from the National Institute of Standards and Technology (NIST) \citep{standard_kyber} that resist the potential threats in FL systems.     
    We utilized ratcheting mechanisms for each training round that provide forward secrecy and post-compromise security for the update models.     
    \item We also propose a hybrid communication approach that combines blockchain transactions and network communication (off-chain and on-chain channels) for transfer models, which enhances security layers and overcomes on-chain costs.   
    \item PQBFL uses blockchain in two aspects: first, as a key establishment facilitator and second, as a decentralized ledger to record the transactions of participants and servers in the FL system.     
    Moreover, we illustrate the capability of blockchain in authentication based on a reputation mechanism and its effectiveness in preserving the privacy of participants in FL systems. \end{enumerate}  
The remainder of this paper is organized as follows. 
Related works are described in Section \ref{Related Work}.  In Section \ref{Preliminaries}, we provide preliminary concepts, including post-quantum cryptography primitives and the structure of the federated learning frameworks.  
Section \ref{PQBFL Protocol} first presents the security and privacy requirements and then describes the PQBFL protocol in detail.  
In Section \ref{Security Analysis}, we present the security analysis and discuss the security capabilities of the proposed scheme.  
Section \ref{Performance Analysis} evaluates the proposed protocol and performs experiments on computation and communication costs. 
Finally, in Section \ref{Conclusion and Future Work}, we present the conclusions of this study and outline future directions for enhancing data privacy in the PQBFL.

%===================================
\section{Related Work}
\label{Related Work}
%===================================

Academics and industries across multiple domains are actively exploring post-quantum cryptographic solutions in response to the emerging threats posed by quantum computing advancements.
In the realm of end-to-end secure messaging applications, Signal Messenger recently published a new version of their Extended Triple Diffie-Hellman (X3DH) protocol, called PQXDH, as a quantum-secure protocol \citep{Signal_pqxdh}. 
Although PQXDH uses the NIST standard Key Encapsulation Mechanism (KEM), Kyber \citep{kyber}, which provides post-quantum forward secrecy and a form of cryptographic deniability, it still relies on the hardness of the discrete log problem for authentication. 
Moreover, Apple proposed a post-quantum security protocol called PQ3 for conversations in the iMessage application using Kyber, which was made available to the public with iOS 17.4 and macOS 14.4 \citep{Apple_pq3}.  
 Similarly, the Transport Layer Security (TLS) protocol \citep{TLS}, which secures communication between web browsers and servers, is undergoing a transformation to mitigate quantum threats.  
 The Open Quantum Safe (OQS) \citep{OQS_project} is an open-source project that aims to support the transition to quantum-resistant cryptography.  
 They integrated a library called liboqs into the forks of BoringSSL and OpenSSL1.1.1, and a standalone OQS provider for OpenSSL3 to provide a prototype post-quantum key exchange, authentication, and ciphersuites in a hybrid key exchange in TLS 1.3 \citep{hybrid_key_exchange_TLS_1.3}.  
 
 In the field of federated learning, researchers seek to provide a security mechanism that guarantees the privacy of participants in collaboration with FL projects using various approaches like Homomorphic Encryption (HE) and Differential Privacy (DP) \citep{A_survey_on_security}. 
 As Table \ref{tab: Comparison } shows, only two studies, LaF \citep{Performance_Analysis} and BFL\citep{LaF}, considered post-quantum security concerns in FL environments.  
 This table compares recent studies and identifies their contributions. 
 For instance, Gurung et al. \citep{Performance_Analysis} combined two post-quantum signature schemes, Dilithium and XMSS, to sign transactions in blockchain-based FL.  
 In this study, participants transfer models through signed transactions, which can incur blockchain costs for information transfers.  
 This scheme can provide post-quantum authentication for transmission models; however, the confidentiality of the update models has not been addressed.   Other quantum security studies in the field of federated learning include those of \citep{LaF} and \citep{A_post-quantum_secure}.   
 These two schemes are improved versions of the Google Group scheme \citep{Practical_secure_aggregation} that uses secret sharing to increase the prevention of privacy models against honest-but-curious servers. 
 In these studies, the authors employed two lattice-based cryptosystems, NewHope \citep{NewHope} and Kyber, to encrypt shares between the server and participants. 
 Given that FL systems typically involve several rounds of training, these schemes require key exchange for each round.   
 Although this can provide forward and post-compromise secrecy, it creates heavy data, communication overhead, and time consumption because of the post-quantum key size, which is usually much larger than traditional ones.

\begin{table}[ht]
\large
\centering
\caption{Comparison functionality of related schemes}
\label{tab: Comparison }
\begin{adjustbox}{width=3.2in}
\begin{tblr}{
  column{even} = {c},
  column{3} = {c},
  column{5} = {c},
  column{7} = {c},
  vline{2} = {-}{},
  hline{1-2,11} = {-}{},
}
Capability            &  DAFL  &   BSAFL &  BESIFL   & LaF & BFL  & PQBFL \\
Decentralization      &  \checkmark   &  \checkmark   &   \checkmark   &  \ding{53}   &  \checkmark  & \checkmark   \\
Authentication        &  \checkmark   &  \checkmark   &   \checkmark   &  \ding{53}   &  \checkmark  & \checkmark   \\
Traceability          &  \checkmark   &  \checkmark   &   \checkmark   &  \ding{53}   &  \checkmark  & \checkmark   \\
User privacy          &  \checkmark   &  \ding{53}    &   \checkmark   &  \ding{53}   &  \ding{53}   & \checkmark   \\
Confidentiality       &  \ding{53}    &  \checkmark   &   \ding{53}    &  \checkmark  &  \ding{53}   & \checkmark   \\
Quantum-security      &  \ding{53}    &  \ding{53}    &   \ding{53}    &  \checkmark  &  \checkmark  & \checkmark   \\
Lightweight           &  \checkmark   &  \ding{53}    &   \ding{53}    &  \ding{53}   &  \ding{53}   & \checkmark   \\
Forward secrecy       &  \ding{53}    &  \ding{53}    &   \ding{53}    &  \checkmark  &  \ding{53}   & \checkmark   \\
Post-compromise       &  \ding{53}    &  \ding{53}    &   \ding{53}    &  \checkmark  &  \ding{53}   & \checkmark  
\end{tblr}
\end{adjustbox}
\end{table}

In addition, \cite{Blockchain-based_decentralized} proposed DAFL as a lightweight digital signature method that facilitates batch verification for authentication to provide a decentralized and simpler framework for FL authentication. 
\cite{Blockchain_and_signcryption} considered the problem of centralized single-layer aggregation in FL and proposed a distributed aggregation architecture called BSAFL by integrating blockchain. 
They introduced signcryption schemes to guarantee the authenticity and confidentiality  of messages in the FL. 
These schemes use quantum-vulnerable cryptography methods to verify the identities of parties. 
They also share public key information and models through blockchain transactions, which are extremely expensive and not cost-effective in the real world. 
Moreover, \cite{BESIFL} proposed a BESIFL paradigm for distributed environments, such as IoT, which leverages blockchain to achieve security using a fully decentralized FL system, integrating mechanisms for the detection of malicious nodes and incentive management in a unified framework.

The PQBFL does not utilize blockchain for key exchanges or model transfers; instead, it uses it to improve security, decentralization, and tracking keys and models.  
In addition to quantum security, the proposed scheme employs a key ratchet mechanism that eliminates the need to exchange keys during each training round.  
It can reduce network overhead and provide forward and post-compromise security, which is ideal for FL systems to improve their performance and security.  
These claimed advantages of the PQBFL are demonstrated and discussed in detail throughout the paper.

%============================
\section{Preliminaries}
\label{Preliminaries}
%============================

We organize the preliminary concepts in this section as follows: In Section \ref{Federated Learning}, we explain federated learning concepts.
In Section \ref{Cryptographic Algorithms}, we address the cryptographic primitives utilized in the PQBFL protocol.

\subsection{Federated Learning}
\label{Federated Learning}
\begin{figure}[ht]
\centering
\includegraphics[width=3in,height=2.2in]{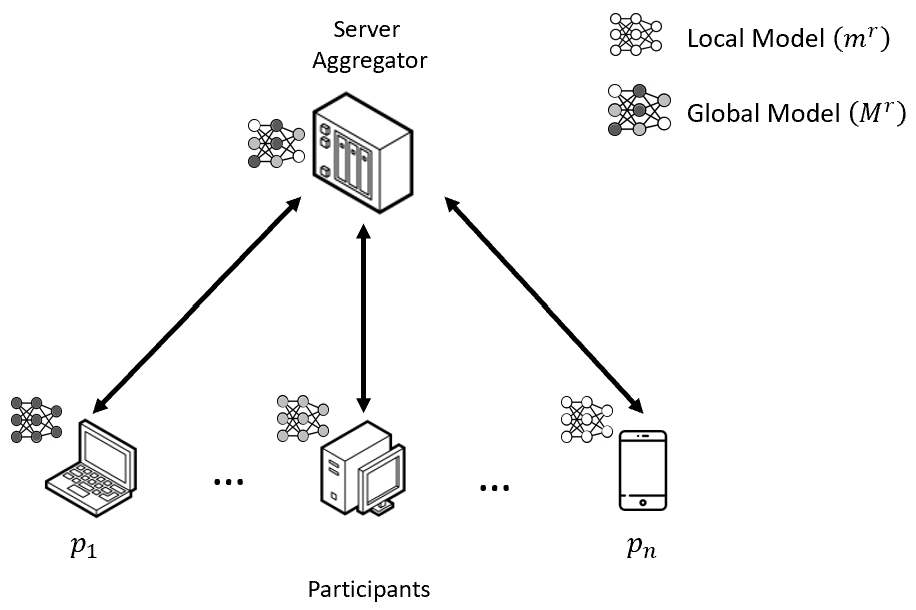}
\caption{Federated learning architecture}
\label{fig:FL architecture}
\end{figure}

%------------------------------
Conventional machine learning typically relies on a centralized data approach, in which data owners upload datasets.
By contrast, the FL mechanism presents a decentralized training method that prioritizes privacy based on distributed data \citep{A_survey_on_federated}.
In FL, participant $p_{i} \subseteq P$ where $i \in [1,n]$ connects to a central server $S$ and contributes to the training task by sending local model update instead of raw data to the server in several rounds, as shown in Figure \ref{fig:FL architecture}.
This can reduce communication overhead, in addition to preserving the privacy of the client participants data.

During each training round, the FL server dispatches the initial model updates to a subset of FL participants.
Each FL participant $p_{i}$ in round $r$ trains its local model $m_{i}^{r}$ using their specific datasets, and then sends the local model update to the FL server, where it gathers and aggregates them to create the joint global model using the following equation:

\begin{equation}
    %M^{r+1}=M^{r}+\eta \frac{\sum_{i=1}^{n^{r}} {w_{i}m_{i}^{r}}}{\sum_{i=1}^{n^{r}}{w_{i}}}
    M^{r}=\sum_{i=1}^{N} \frac{{w_{i}}}{N} m_{i}^{r}
\end{equation}

Here, $M^{r}$ denotes the global model in the $r$-th round, where $r>0$. 
In addition, $m_{i}^{r}$ denotes the local model of $i$-th FL participant, $w_{i}$ and $N$ are the weights applied to $m_{i}^{r}$ values and number of participants, respectively.
Although FL increases privacy and efficiency, there are concerns regarding different types of attacks such as authentication, model secrecy, free-riding, and single-point failure, which we discuss in Section \ref{Security and Privacy requirements}.

\subsection{Cryptographic Algorithms}
\label{Cryptographic Algorithms}
%------------------------------
This section discusses traditional cryptographic algorithms and post-quantum primitives recently adopted as part of the NIST standard \citep{standard_kyber}, which are integrated into the PQBFL design. 
These algorithms include key exchange, encapsulation mechanisms, key derivation, and digital signatures, all working together to establish the security foundation of the proposed scheme.

\begin{comment}
    
\subsubsection{Signature}

SPHINCS$^+$ is a stateless, hash-based signature standard used in security protocols \cite{sphincs_site}. 
It instantiates using three different hash algorithms: SHA-256, SHAKE256, and Haraka. 
This method offers small public and private keys of almost the same size as the keys of the traditional ECDSA algorithm, but it provides significant signature  sizes up to that amount depending on the parameter selections. 
The small size of SPHINCS$^+$ keys, particularly the public-key size, comparable to other quantum-secure signatures such as Dilithum[] and Falcon[], renders it highly suitable for distribution via blockchain networks. 
SPHINCS$^+$ signature, similar to traditional algorithms, consists of the following algorithms:

\begin{itemize}
    \item $\left ( sk ,pk\right )\leftarrow sphincs.KeyGen\left (  \right )$:  It uses secret seed generate all the private key $sk$ an public key $pk$. 
    \item $S\leftarrow sphincs.Sign\left ( m,sk \right )$: The sign algorithm employ secret key $sk$ and message $m$ as input and the output is a signature $S$ which is send along with $m$.
    \item $\{0,1\}\leftarrow sphincs.verify\left ( m,S,pk \right )$: One can verify that a signature $S$ is valid for a message $m$ if and only if the output is $1$ otherwise $0$.
\end{itemize}

\end{comment}

\subsubsection{Signature}
Digital signatures authenticate the received models and validate the transactions on the blockchain. 
Currently, popular blockchains such as Ethereum \citep{Ethereum} continue to rely on conventional signatures because there are no concerns regarding Harvest-Now, Decrypt-Later attacks on signature algorithms and the existence of significantly powerful quantum computers \citep{Google_blog}.
Consequently, the PQBFL protocol relies on the standard Elliptic Curve Digital Signature Algorithm (ECDSA) for both the integrity and authenticity of on- and off-chain communication.
The key generation algorithm $(sk,pk) \leftarrow Sig.Gen(k)$ receives a random security parameter $k$ as input and generates a secret key $sk$ and a public key $pk$; the signing procedure $ \sigma \leftarrow Sig.Sign(sk,m)$ requires the secret key $sk$ and  message $m$ as input and generate a signature $\sigma$, and  the verification algorithm $\{0,1\} \leftarrow Sig.Ver(pk,m,\sigma)$ takes as input the public key, message and signature, and returns a bit to indicate the validity of the signature. 
To guarantee correctness, we need $\forall m$

\begin{equation}
    \begin{split}
        \Pr [(sk,pk)\leftarrow Sig.Gen(k),\sigma \leftarrow Sig.Sign(sk,m):\\ 1 \leftarrow Sig.Ver(pk,m,\sigma) ] =1  
    \end{split}
\end{equation}

The security requirement for a digital signature $\sigma$ is that it is unforgeable under a chosen message attack (EUF-CMA). Thus, it is infeasible for an attacker to generate a new verifiable message and signature pair, even with access to a signing oracle. more precise enough:
\begin{equation}
    \text{Adv}_{\sigma}^{\text{EUC-CMA}}(\mathcal{A}) = \Pr \left[ \text{EUC-CMA}_{\sigma}^{\mathcal{A}} \rightarrow 1 \right] 
\end{equation}

\noindent where $\text{EUC-CMA}_{\sigma}^{\mathcal{A}}$ denotes the security experiment.
These equations ensure the security, integrity, and authentication of a signature scheme, thereby forming a critical component of the proposed protocol.

\subsubsection{KDF}
Key Derivation Functions (KDF) are  cryptographic algorithms that uses a pseudo-random function $PRF$ to derive secret keys from a secret value, such as a root key. 
HKDF \citep{Cryptographic_extraction} is a simple KDF designed based on the extract-then-expand procedure using the message authentication code, HMAC. 
This logically consists of two sequential algorithms:
\begin{enumerate}
    \item $k_{e} \leftarrow HKDF.Extract(rk, s)$: A deterministic extraction algorithm that generates an extract key $k_{e}$ as the output after receiving the root key material $rk$ and salt $s$ as inputs.
    \item $k^{*} \leftarrow HKDF.Expand(k_{e}, label, l)$: This is also a deterministic expansion algorithm in which its input is a key $k_{e}$ (output of the $Extract$ algorithm), a label $label \in \{0,1\}^{*}$, and length parameter $l$, which finally outputs a binary key $k^{*}$ with length $l$.
\end{enumerate}

Given that the security of the KDF originates from the security of  $PRF$, these algorithms are considered secure if it is computationally impossible for an attacker to discern the output of a pseudo-random function from a truly random function. 
In other words:

\begin{equation}
    \begin{aligned}
        \text{Adv}_{F}^{\text{prf}}(\mathcal{A}) = \Big| \Pr \left[ \mathcal{A}^{F(k, \cdot)} = 1 \mid k \leftarrow \mathcal{K} \right] - \\ 
        \Pr \left[ \mathcal{A}^{R(\cdot)} = 1 \mid R \leftarrow \mathcal{R} \right] \Big| \leq \epsilon 
    \end{aligned}
\end{equation}

This guarantees the security requirements of the ratcheting mechanism of the PQBFL protocol.

\subsubsection{Diffie-Hellman Key exchange}
The Elliptic Curve Diffie-Hellman (ECDH) algorithm is a traditional key exchange cryptographic protocol used to establish a shared secret between two parties.
It typically operates on a curve in the form of $y^2=x^3+ax+b\:(\!\!\!\!\mod {p})$ where $(p,a,b,G)$ are public domain parameters. 
ECDH is based on the following property of EC points: Each party computes its respective public key points $A=(k_{a} * G)$ and $B=(k_{b} * G)$, leveraging its private key numbers $k_{a}$ and $k_{b}$ (e.g., Alice and Bob).
Once these points are exchanged, the shared secret $ss$ is derived for each party using the other party's public key and private key through the following operation:

\begin{equation}
    A * k_{b} = B * k_{a}= ss
\end{equation}

This shared secret generated from ECDH, together with the shared secret of KEM, is used to generate the encryption keys of the models in our FL protocol.

\subsubsection{Key Encapsulation Mechanism}
To establish a secure channel between the participants and the server in the FL to transferring models, we used a post-quantum secure key encapsulation mechanism.
Kyber \citep{kyber} is a lattice-based and IND-CCA2-secure KEM that relies on the difficulty of solving the learning with errors in the module (M-LWE) problem. Kyber KEM involve the following algorithms:

\begin{itemize}
    \item $\left ( sk ,pk\right )\leftarrow Kyber.KeyGen\left (  \right )$: It uses a secret randomized seed to generate private key $sk$ and public key $pk$.
    \item $ (ct,ss)\leftarrow Kyber.Encap\left ( pk \right )$: The encapsulation algorithm employs the public key $pk$ as the input, and the output is a ciphertext $ct \in \mathcal{C}$ and a unique shared secret key $ss \in \mathcal{K}$  used for subsequent encryption and decryption.
    \item $ss \leftarrow Kyber.Decap (sk,ct)$: the ciphertext $ct$ and private key $sk$ are employed to generate the same shared secret key $ss$ on the opposite side of the channel for subsequent encryption and decryption.
\end{itemize}
If an adversary $\mathcal{A}$ cannot computationally identify the shared secret $ss$ of a challenge ciphertext $ct$ from a random shared secret, even with access to a decapsulation oracle, then the KEM is considered secure under Indistinguishable under Chosen Ciphertext Attack (IND-CCA). 
This security requirement is formalized as follows:

\begin{equation}
    \begin{aligned}
        \text{Adv}_{\Pi}^{\text{ind-cca}}(\mathcal{A}) = \Big| \Pr \left[ \text{IND-CCA}_{\Pi}^{\mathcal{A}} \rightarrow 1 \right] - \frac{1}{2} \Big| 
    \end{aligned}
\end{equation}

\noindent where $\text{IND-CCA}_{\pi}^{\mathcal{A}}$ denotes the security experiment. 
The correctness requirement for a two-party system is also defined such that the shared secrets $ss$ on both sides must be identical.

%============================
\section{PQBFL Protocol}
\label{PQBFL Protocol}
%============================
In this section, we first examine the security and privacy requirements of the PQBFL, and then present the system model. 
We assumed that the server and participants in the federated learning system operate in a zero-trust environment based on a reputation mechanism. 
First, all participants in the system are considered semi-honest, achieving rewards and punishments over time based on their performance.

\subsection{Security and Privacy requirements}
\label{Security and Privacy requirements}
%------------------------------
 
The main security and privacy requirements of the proposed protocol are as follows.

\noindent\textbf{Authentication.}
The first step in a federated learning network is mutual authentication between the server and participants to prevent security threats, such as man-in-the-middle (MITM) attacks.
Participants must be able to verify the global models sent by the server based on the registered project and published tasks transactions on the blockchain.
In addition, the server must verify the identities of the project participants in different rounds based on their registration information. 
However, it is possible that an authenticated participant intentionally or unintentionally sends poisoned data or model to the aggregator server. 
Thus, authentication alone is not the criterion for a participant to be honest, demonstrating the need for a reputation-based mechanism.

\noindent\textbf{Confidentiality.}
In FL, the confidentiality of the models exchanged between the server and participants is crucial.
In the absence of secrecy, adversaries can capture local models and employ them in subsequent attacks, such as Membership Inference Attacks (MIA) \citep{MIA_attack} and Source Inference Attacks (SIA) \citep{SIA_attack}. 
In addition, an attacker can collect local model updates transmitted between the FL server and the participants, allowing a free-ride attack with free-obtained model updates without contributions. 
PQBFL has inspired the ratcheting key technique, which is utilized in the PQXDH \citep{Signal_pqxdh} and PQ3 \citep{Apple_pq3} messaging protocols, to guarantee model confidentiality and provide post-compromise security and forward secrecy of transmitted models in different rounds.
Forward secrecy protects the model against a potential compromise from the previous exchange model, thus ensuring the confidentiality of the previously exchanged models. 
Post-compromise security ensures the security of future exchange models if current keys are compromised.

\noindent\textbf{Replay protection.}
Although the authentication and secrecy of local models prevents impersonating participants and the disclosure of model specifics in an FL system, adversaries can eavesdrop on the channel between the server and participants and intercept model updates in each round. 
Subsequently, they can replay the previously transmitted local models in subsequent training rounds. 
This attack disrupts the continuous updating of the global model with fresh local models, leading to a reduction in overall performance. PQBFL aims to provide cryptographic and blockchain-based replay protection mechanisms to prevent such attacks.

\noindent\textbf{Quantum-security.}
Currently, a pressing concern regarding quantum computers is passive quantum attacks such as HNDL,  because quantum computers are still in their infancy and do not pose a threat to the creation of active quantum attacks.
This renders the utilization of traditional signature schemes reasonable for existing blockchains \citep{Google_blog}.
The PQBFL can prevent passive quantum attacks, and once post-quantum blockchains become available, it can easily adopt them and resist active quantum attacks.
Moreover, we used traditional schemes in a hybrid manner because we cannot completely rely on post-quantum schemes due to insufficient research and potential vulnerabilities \citep{An_efficient_key,Transition_to_Post-Quantum}.

\noindent\textbf{Privacy preserving.}
Privacy preservation for data and participants is one major concern with FL systems. 
Nonetheless, the existence of a curious-but-honest server and threats such as the MIA and SIA model updates jeopardize participants and data privacy.
Blockchain can be an ideal solution to provide a balance between the participant's pseudo-anonymity to increase privacy and tracking malicious participants who attempt to send a suspicious model. 
However, model data privacy was beyond the scope of this study.

\noindent\textbf{Single-point failure mitigation.}
FL depends heavily on the central server, leading to a single-point failure vulnerability stemming from potential Distributed Denial-of-Service (DDoS) attacks on the server, as shown in Figure \ref{fig:FL architecture}.  
To address this concern, the PQBFL employs  blockchains to mitigate such vulnerabilities.  This approach requires participants to execute transactions with a low transaction fee, which minimizes the transmission of fake update models.

\subsection{System model}
\begin{figure}[ht]
    \centering
\includegraphics[width=3.2in,height=2.7in]{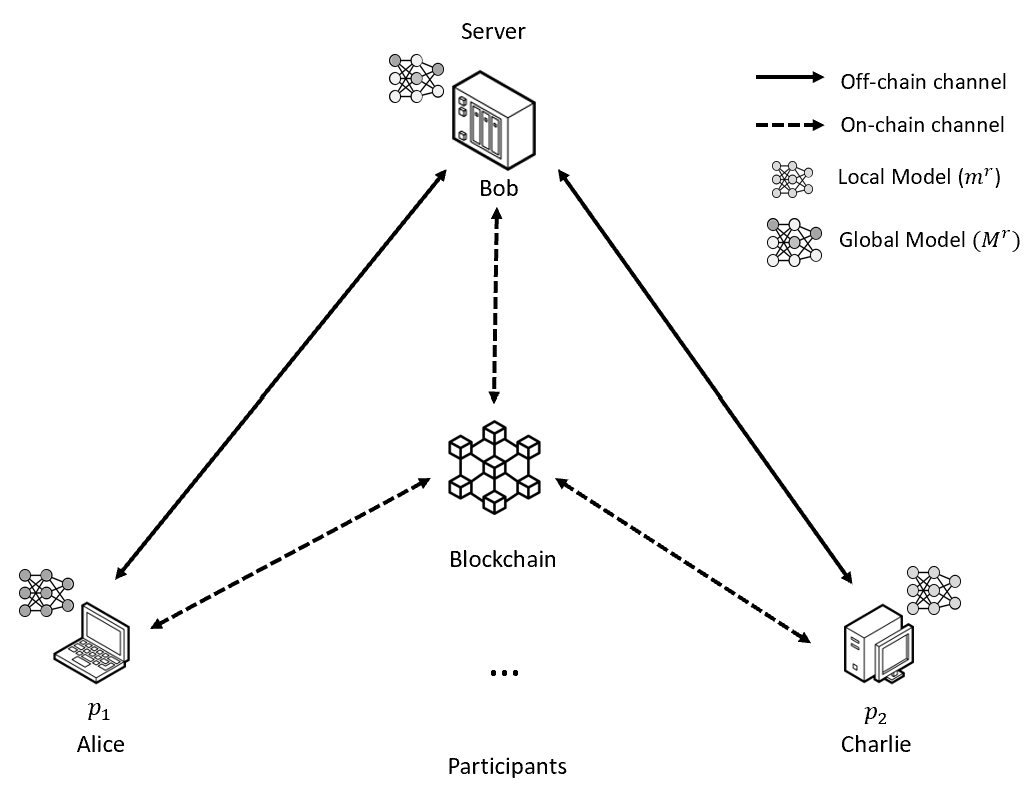}
    \caption{The proposed federated learning architecture}
    \label{fig:proposed FL architecture}
\end{figure}

PQBFL benefits from symmetric and asymmetric key ratcheting for each training round to retain efficiency and security against adversary access to a quantum computer.
In this section, we explain the PQBFL protocol, which comprises three primary components: the blockchain, server, and the participants. 
The blockchain plays a pivotal role, initially serving as a key establishment facilitator to securely set cryptographic keys between participants and the server.
In addition, it functions as a decentralized ledger that records all transactions, including project registration, task publishing, model updates, and feedback, within the FL framework.
Although participants and server transmit models securely on the off-chain channel, they also send their corresponding transactions, as shown in Figure \ref{fig:proposed FL architecture}, on the on-chain channel.
Therefore, this approach enables pseudo-anonymous tracking, rewarding, and penalizing participants and increases the efficiency of blockchain transaction costs.
The next subsection elaborates on the PQBFL procedure using Alice as a participant and Bob as the aggregator server with the notations stated in Table \ref{tab: Notations}.

%\begin{figure}[ht]
%    \centering
%\includegraphics[width=3.2in,height=3in]{Images/2.png}
%    \caption{List of notations}
%    \label{fig:Notations}
%\end{figure}


\begin{table}
\small
\centering
\caption{List of notations}
\label{tab: Notations}
\begin{adjustbox}{width=3.2in}
\begin{tabular}{l|l}
\hline
\rowcolor[HTML]{EFEFEF} 
Symbol                 & Description                                                 \\\hline
($ksk_{b},kpk_{b}$)    & Private and public KEM keys of Bob                          \\[0.8ex]
$SS_{k}$               & KEM Shared secret                                           \\[0.8ex] 
($esk_{b},epk_{b}$)    & Private and public ECDH keys of Bob                         \\[0.8ex]
$SS_{e}$               & ECDH Shared secret                                          \\[0.8ex]
$(ssk_{b},spk_{b})$    & ECDSA blockchain signature key pair                         \\[0.8ex]
$L_{j}$                & Symmetric ratcheting threshold in $j$-th asymmetric ratchet             \\[0.8ex]
$KDF_{S},KDF_{A}$      & Symmetric and asymmetric key derivation functions           \\[0.8ex]
$RK_{j}$               & Root key at $j$-th asymmetric ratchet                       \\[0.8ex]
$CK_{ij}$              & Chain key of $i$-th symmetric in $j$-th asymmetric ratchet  \\[0.8ex]
$K_{ij}$               & Model key of $i$-th symmetric in $j$-th asymmetric ratchet  \\[0.8ex]
$r$                    & The round number                                            \\[0.8ex]
$T$                    & Terminate task trigger                                      \\[0.8ex]
$D_{t}$                & Deadline task                                               \\[0.8ex]
$id_{p},id_{t}$        & The project and task identifiers                             \\[0.8ex]
$m^{r},M^{r}$          & Local and Global Model at $r$-th round                      \\[0.8ex]
$h(x)$                 & One-way hash functions                                      \\ \hline
\end{tabular}
\end{adjustbox}
\end{table}

\subsubsection{Registrations}
We assume that Bob aims to initiate a new FL project that involves Alice's participation. 
Bob has already deployed a smart contract on the blockchain that governs the project lifecycle, as outlined in Algorithm \ref{alg:Smart Contract}. 
This smart contract ensures authentication, security of keys, transparency and fairness by requiring a deposit from the project initiator, serving as a guarantee to aggregate local models and reward participants. 
It leverages event-driven mechanisms to manage essential operations such as registering clients and projects, publishing tasks, handling model updates, providing feedback, and finalizing the project.
This structure ensures accountability, secure interactions, and automated enforcement of agreements, fostering trust among all participants.
Then, the session establishment and training process begins with the following steps:

{\footnotesize
\begin{algorithm}
\caption{PQBFL Smart Contract}
\label{alg:Smart Contract}
%\SetKwInput{KwInput}{Input}                % Set the Input
%\SetKwInput{KwOutput}{Output}              % Set the Output
\DontPrintSemicolon

\textbf{Event} \texttt{\small{RegClient}}($cAddr$, $id_p$, $sc$, $h\_epk$ );\\
\textbf{Event} \texttt{\small{RegProject}}($id_{p}$, $nClients$, $sAddr$, $h\_M^{0}$, $h\_pks$);\\
\textbf{Event} \texttt{\small{Task}}($r$, $h\_M^{r}$, $h\_pks^{r}$, $id_{p}$, $id_{t}$, $nClients$, $D_{t}$, $time$);\\
\textbf{Event} \texttt{\small{Update}}($r$, $h\_m^{r}$, $h\_c\_epk$, $id_{p}$, $id_{t}$, $cAddr$, $time$);\\
\textbf{Event} \texttt{\small{Feedback}}($r$, $id_{p}$, $id_{t}$, $h\_m^{r}$, $h\_pks^{r}$, $cAddr$, $sc$, $T$);\\
\textbf{Event} \texttt{\small{ProjectTerminate}} ($r$, $id_{p}$, $id_{t}$, $time$);\\

% Set Function Names
\SetKwFunction{FRegisterClient}{RegisterClient}
\SetKwFunction{FRegisterProject}{RegisterProject}
\SetKwFunction{FPublishTask}{PublishTask}
\SetKwFunction{FUpdateModel}{UpdateModel}
\SetKwFunction{FFeedbackModel}{FeedbackModel}
\SetKwFunction{FUpdateScore}{UpdateScore}
\SetKwFunction{FFinishProject}{FinishProject}
%\SetKwFunction{FFinishTask}{FinishTask}
%\SetKwFunction{FIsTaskDone}{IsTaskDone}
\SetKwProg{Fn}{Func}{:}{\KwRet}
% RegisterProject Function
\Fn{\FRegisterProject{$id_{p}$, $n$, $h\_M^{0}$, $h\_pk$}}{
    $nClients \gets n$\\
    $sAddr \gets msg.sender$\\
    \If{$Deposit < sAddr.value$ \textbf{and} $Done[id_{p}] = False$}{    
        $projects[id_{p}] \gets \{id_{p}, nClients, sAddr, \\
        times, h\_M^{0}, h\_pk_{b}\}$\;
    }
Emit event \texttt{RegProject}\;
}
\;

% Function Definitions
\SetKwProg{Fn}{Func}{:}{\KwRet}
% RegisterClient Function
\Fn{\FRegisterClient{$h(epk)$, $id_{p}$}}{
    \If{$project[id_{p}].Clients < nClients$}{
        $cAddr \gets msg.sender$\;\\
        $h\_epk \gets h(epk)$\;\\
        $sc \gets 0$\;\\
        $clients[Addr] \gets \{cAddr, id_{p}, sc, h\_epk\}$\\
        $project[id_{p}].clients \gets project[id_{p}].clients + 1$}\;
Emit event \texttt{RegClient}\;
}

\;

\SetKwProg{Fn}{Func}{:}{\KwRet}
% PublishTask Function
\Fn{\FPublishTask{$r$, $h\_M$, $h\_pks$, $id_{t}$, $id_{p}$, $D_{t}$}}{
    $sAddr \gets msg.sender$\\
    $tasks[id_{t}] \gets \{r, h\_M^{r}, id_{t}, sAddr, h\_pks^{r}, id_{p}, D_{t}, time\}$\;\\
    Emit event: \texttt{Task}\;
}
\;

\SetKwProg{Fn}{Func}{:}{\KwRet}
% UpdateModel Function
\Fn{\FUpdateModel{$r$, $h\_m^{r}$, $h(ct||epk)$, $id_{t}$, $id_{p}$}}{
    \If{$h\_pk \neq None$}{
        $h\_c\_epk \gets h(ct||epk)$;
    }
    \If{$tasks[id_{t}] \neq 0$ \textbf{and} $h\_m^{r} \neq \emptyset$}{
        $Update[id_{t}] \gets \{r, id_{t}, sAddr, h\_m^{r}, h\_c\_epk_,\\ id_{p}, time\}$\; 
    }
    Emit event: \texttt{Update}\;
}
\;

\SetKwProg{Fn}{Func}{:}{\KwRet}
% ProvideFeedback Function
\Fn{\FFeedbackModel{$r$, $id_{t}$, $id_{p}$, $cAddr$, $sc$, $T$}}{
    \If{$tasks[id_{t}] \neq 0$}{
        $Feedbacks[id_{t}] \gets \{r, id_{t}, id_{p}, cAddr, time, sc, T\}$\;\\
        \(UpdateScore\)\{$cAddr$, $sc$\}\; 
    }
    Emit event: \texttt{Feedback}\;
}
\;

\SetKwProg{Fn}{Func}{:}{\KwRet}
% UpdateClientScore Function
\Fn{\FUpdateScore{$cAddr$, $sc$}}{
    $clients[cAddr].score \gets clients[cAddr].score + sc$\;\\
    \If{$clients[cAddr].score < 0$}{
        $clients[cAddr].score \gets 0$\;
    }
}
\;

\SetKwProg{Fn}{Func}{:}{\KwRet}
\Fn{\FFinishProject{$id_{p}$}}{
    \If{$sAddr=msg.sender$}{
        \If{$projects[id_{p}] \neq 0$ \textbf{and} $Done[id_{p}] = False$}{
            $Done[id_{p}] \gets True$\;   
        }
    }
    Emit event: \texttt{ProjectTerminate}\;
}
\;
\end{algorithm}
}

\noindent\textbf{Server key generation.} 
First, Bob uses $Kyber.KeyGen()$ algorithms to generate KEM keys $(ksk_{b},kpk_{b})$.
He also generates an ECDH key pair $(esk_{b},epk_{b})$ using a random number.
    
\noindent\textbf{Server project registration.} 
Bob initiates this process by creating a \textit{Registration} transaction on the blockchain. 
This transaction includes the hash of the ECDH and Kyber public keys, an initial model hash, the number of participants, and a unique project identifier in the form $Tx_{r}= \{h(kpk_{b}||epk_{b}),n,h(M^{0}),id_{p}\}$.
    
\noindent\textbf{Participant key generation.} 
Alice monitors events on the blockchain. 
Upon detecting the transaction $Tx_{r}$, she also generates the ECDH $(esk_{a}, epk_{a})$ key pair.

\noindent\textbf{Participant project registration.} 
Alice decides to register for the project by creating her own \textit{Registration} transaction on the blockchain.
In this transaction, she includes the desired project identifier $id_p$ and hash of the ECDH public key. 
We denote this transaction by $Tx_{r}=\{ h(epk_{a}),id_{p}\}$. 

Now, as shown in the first two steps of Figure \ref{fig:sequence diagram key establishment}, Bob and Alice have the hash of concatenated public keys of each other and can pursue establishing the first root key $RK_{1}$ in the next steps.

\subsubsection{Session establishment}
Session establishment is the second phase, after Bob and Alice’s registration. Upon receiving the registration event of participant on the blockchain, the process begins as follows.

\noindent\textbf{Send Keys.} 
Bob initiates the process by wrapping a message $msg_{b}: <kpk_{b},epk_{b},Tx_{r},id_{p}>$ including his Kyber and ECDH public keys, his registration transaction $Tx_{r}$, along with the project identifier $id_{p}$, respectively.
He then utilizes the same ECDSA private key of blockchain $ssk_{b}$ and signs $msg_{b}$ to send the pair $<msg_{b},\sigma_{b}>$ to Alice through an off-chain channel.

\noindent\textbf{Authentication.} 
Alice validates receipt Bob's signature on $msg_{b}$ using Bob's blockchain public key recovered from previous blockchain transaction. 
If the blockchain is verified, she combines $kpk_{b}$ and $epk_{b}$ and compares their respective hashes in the blockchain transaction $h(kpk_{b}||epk_{b})$. 
If they match, Alice authenticates Bob's KEM and the ECDH public keys. 

\begin{figure*}[ht]
    \centering
\includegraphics[width=6.2in,height=4in]{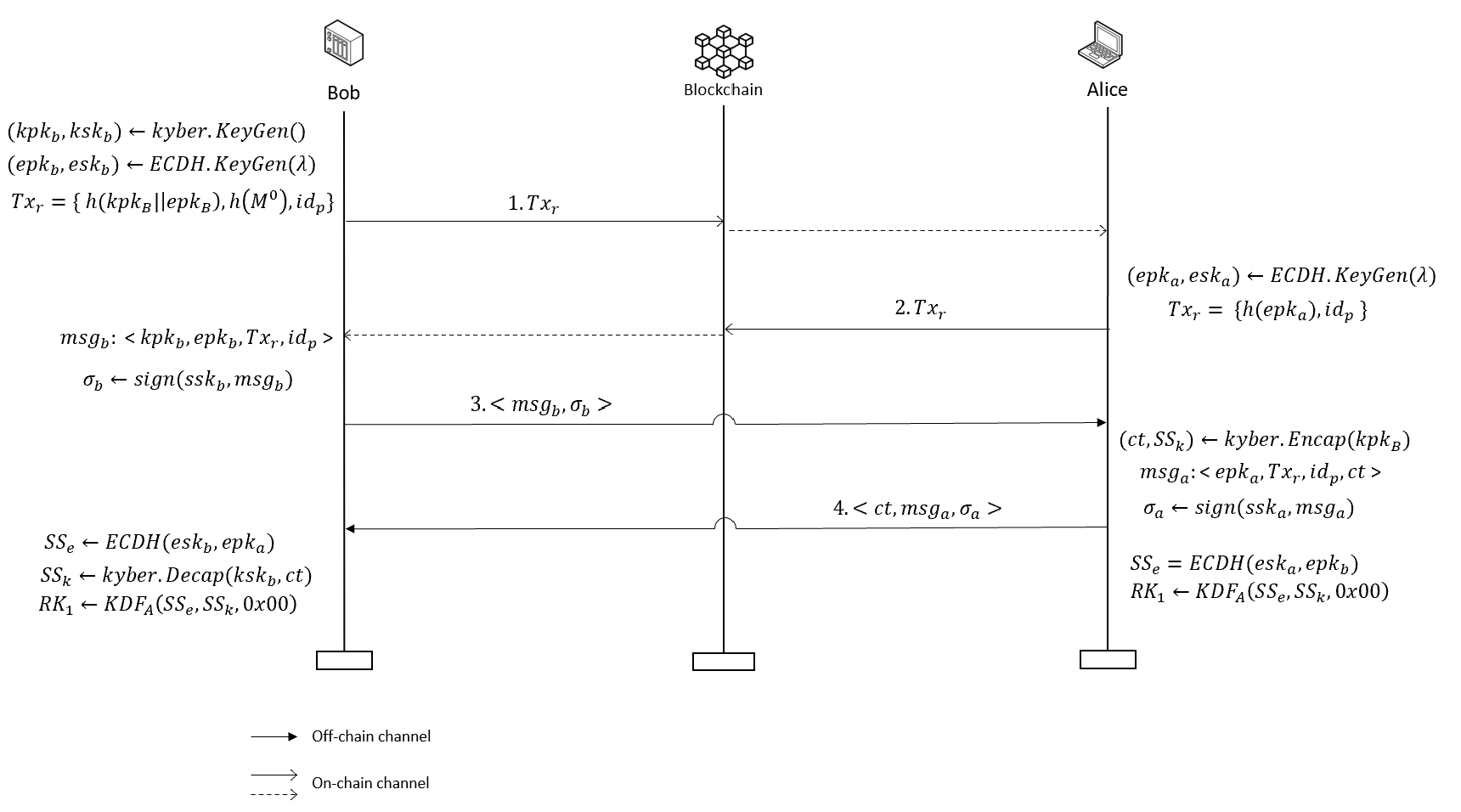}
    \caption{Sequence diagram for registration and session key establishment.}
    \label{fig:sequence diagram key establishment}
\end{figure*}

\noindent\textbf{Generate shared secrets.} 
Alice proceeds to derive the ECDH shared secret $SS_{e}$ using her ECDH private key $esk_{a}$ and Bob's ECDH public key $epk_{b}$. 
Using Bob's KEM public key, Alice obtains the KEM shared secret $SS_{k}$ and cipher-text $ct$ via $Kyber.Encap\left ( kpk_{b} \right )$ function.

\noindent\textbf{Generate root key.} 
Alice uses the KEM and ECDH shared secrets ($SS_{k}$, $SS_{e}$) and a zero byte sequence ($0x00$) as salt and sends them to the asymmetric ratchet $KDF_{A}$ successively, yielding the root key $RK_{j}$ where $j\in [1,n]$ and valid for several training rounds depending on the security level requirement. 

In order to exchange Alice's keys with Bob and establish the corresponding root key for Bob, Alice similarly wrap message $msg_{a}:<epk_{a},Tx_{r},id_{p},ct>$ including the ECDH public key, her registration transaction $Tx_{r}$, project identifier $id_{p}$ and KEM cipher-text $ct$. 
As illustrated in Figure \ref{fig:sequence diagram key establishment}, Alice sends a message and its corresponding signature, $< msg_{a},\sigma_{a}>$ via the off-chain channel. 
Bob then validates the signature and authenticates the received keys using hash of keys. 
He generates an ECDH shared key $SS_{e}$ using the Alice public key $epk_{a}$ and his private key $esk_{b}$.
Furthermore, he provides the received ciphertext $ct$ and his KEM private key $ksk_{b}$ to the $Kyber.Decap()$ function to generate the shared key $SS_{k}$, and both shared secrets pass on $KDF_{A}$ to obtain the corresponding root key $RK_{j}$. 
Now, both parties can proceed and derive the first chain key $CK_{i,j}$ and model key $K_{i,j}$ from their root key.

\subsubsection{Send and receive models}
Alice can begin transmitting her local model when she has completed the training process. 
Model transmission also involves on- and off-chain phases.  

\noindent\textbf{Publish task (Step 1 and 2).} When the root key is established on the server and there are sufficient participants.
Bob (server) broadcasts a \textit{Publish Task} transaction $Tx_{p}$ on blockchain, including round number $r$, project and task identifies, $id_{p}$ and $id_{t}$, hash of wrapped global model information, $h(Inf_{b}^{r})$, and deadline of task $D_{t}$ for participants.
He also generates the chain and model keys, $CK_{i,j}$ and $K_{i,j}$ using $KDF_{S}$ based on the root key $RK_{j}$ for the round number $r$, such that:

\begin{equation}
    r=
    \begin{cases}
        L_{j}\times j+i   &     \text{if $L_{j}$ is fixed} \\[2ex]
    \sum_{j=1}^{n}L_{j}+i   & \text{if $L_{j}$ is varies}
    \end{cases}  
\end{equation}

\noindent  where $L_{j}$ is the number of symmetric ratcheting steps  in $j$-th asymmetric ratchet determined by the server and parameter $i$ shows the number of current symmetric ratchets inside each asymmetric ratchet.
Bob then encrypts $Inf_{b}^{r}$ and $Tx_{p}$ using model key $K_{i,j}$ to generate encrypted message $msg_{b}$. 
Finally, he signs $msg_{b}$ using his blockchain private key, $ssk_{b}$ and sends it to Alice via the off-chain channel.

\begin{figure*}[ht]
    \centering
\includegraphics[width=6.2in,height=4in]{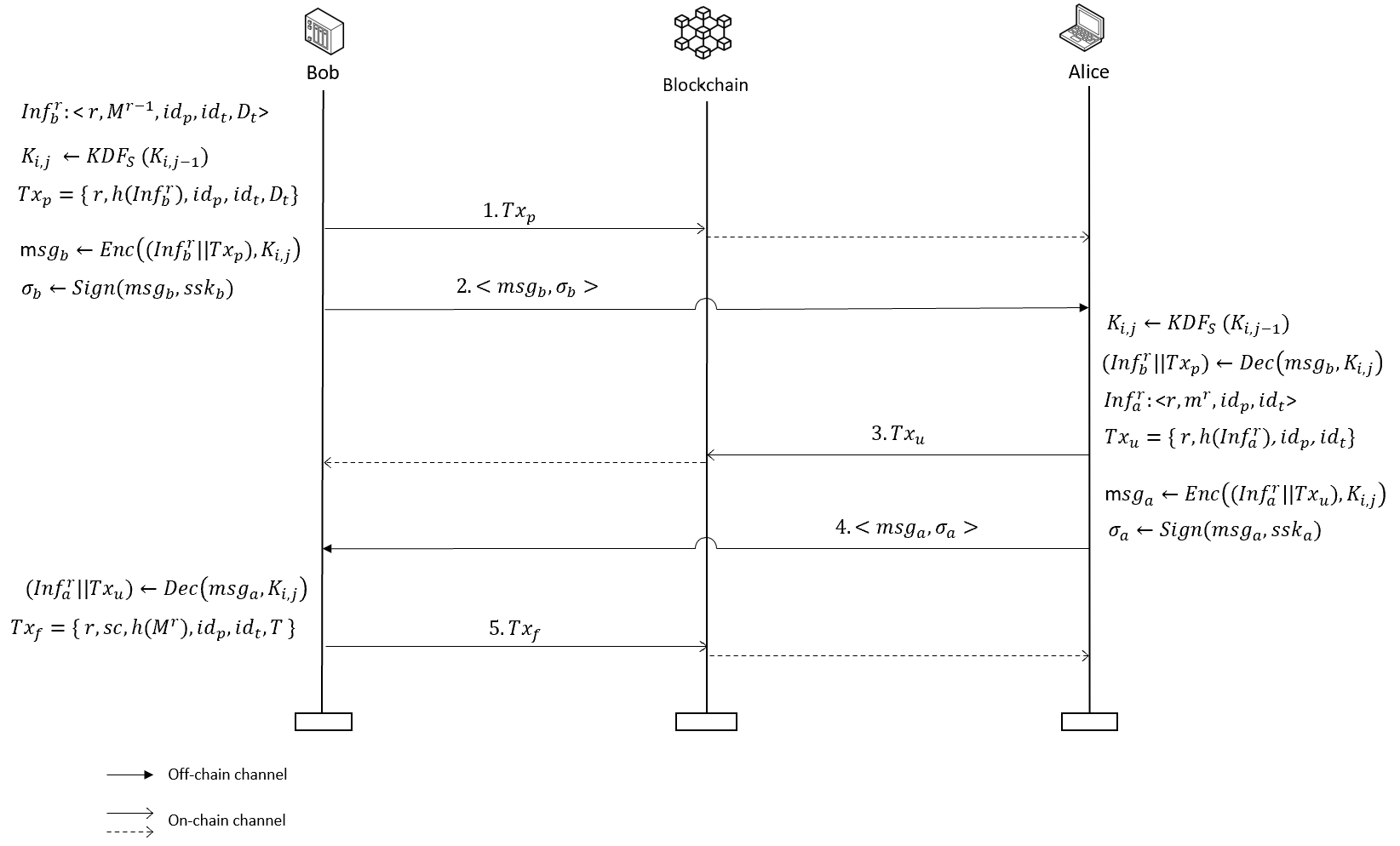}
    \caption{Sequence diagram for the send and receive model in a single training round.}
    \label{fig:sequence diagram send/recive}
\end{figure*}

\noindent\textbf{Update model (Step 3 and 4).}
Alice (participant) receives encrypted information and a signature, first authenticates and decrypts the message $msg_{b}$ using $K_{i,j}$, and then starts training the local model.
After the local model $m^r$ is trained, she makes the \textit{Update Model} transactions $Tx_{u}$ on the blockchain.
This transaction comprises round number $r$, project and task identifies $id_{p}$ and $id_{t}$, and the hash of model information $h(Inf_{a}^{r})$, denoted as $Tx_{u} =\{r, h(Inf_{a}^{r}), id_{p}, id_{t}\}$.
To send an updated local model to Bob, as illustrated in Figure \ref{fig:sequence diagram send/recive}, Alice first encrypts the model information $Inf_{a}^{r}$ and transaction $Tx_{u}$ using the generated model key $K_{i,j}$ where $K_{1,1}=KDF_{A}(RK_{1})$ to construct message $msg_{a}$.
Subsequently, Alice sends signed messages $\sigma_{a}$ and $msg_{a}$ over an off-chain channel to Bob.

\noindent\textbf{Feedback model (Step 5).}
Bob is first aware of the update task information performed by Alice from the blockchain transaction $Tx_{u}$ and then receives the message from the model through the off-chain channel.
Bob authenticates the signature using Alice's blockchain public key.
He decrypts the received $msg_{a}$ using $K_{i,j}$ and sends it to subsequent functions, namely, analyzing the model and determining a reward or penalty for Alice.
If approved, he sends the \textit{Feedback model} transaction $Tx_{f}$, which includes the score $sc$, hash of the global model $M^r$, round number $r$, project identifier $id_{p}$, and task identifier $id_{t}$ to the blockchain and aggregates it with other local models.
This iterative process continues until $T=1$, triggering the termination of the task in the $Tx_{f}$ transaction and the completion of the FL learning process.

\newcommand{\tikzcircle}[2][red,fill=red]{\tikz[baseline=-0.5ex]\draw[#1,radius=#2] (0,0) circle ;}%
\begin{figure*}[ht]
    \centering
\includegraphics[width=6.2in,height=4.8in]{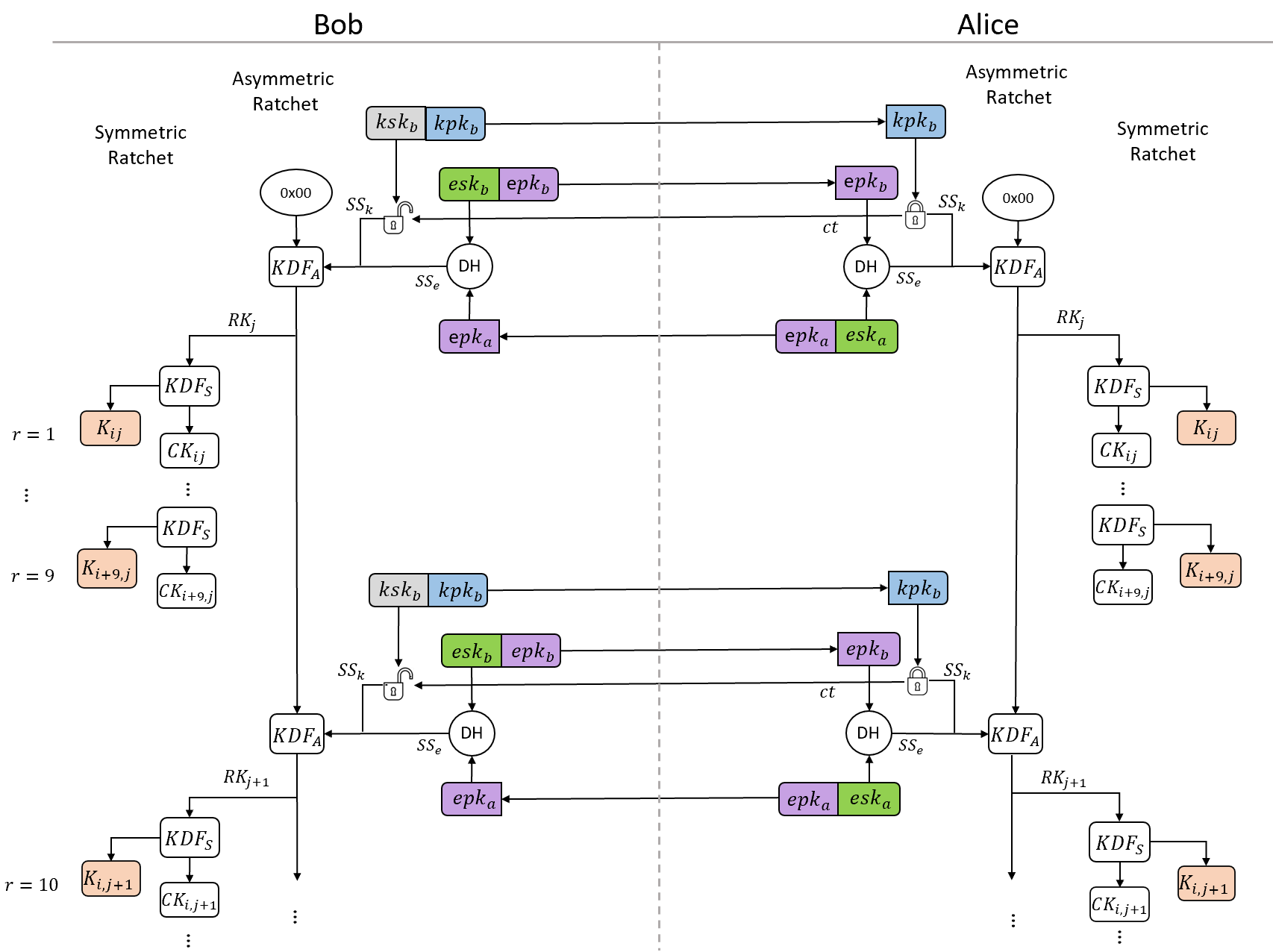}
    \caption{
    Illustration of PQBFL key exchange, encapsulation, decapsulation and derivation processes for different FL training rounds, $r$.
    Initially, Bob and Alice send public keys (\tikzcircle[LimeGreen, fill=LimeGreen]{3.5pt}\tikzcircle[Orchid, fill=Orchid]{3.5pt} ECDH key pairs, \tikzcircle[Gray, fill=Gray]{3.5pt}\tikzcircle[CornflowerBlue, fill=CornflowerBlue]{3.5pt} KEM key pair), which leads to the first asymmetric ratcheting and generation of the first root key $RK_{j}$. 
    Then, given $L_{j}=9$, nine distinct keys (\tikzcircle[Apricot, fill= Apricot]{3.5pt} model key)  are consecutively derived for each model round using symmetric ratcheting, $KDF_{S}$. 
    The second asymmetric ratcheting is trigger by Bob again to derive the model keys for the $10$-th round in Bob and Alice party.}
    \label{fig:PQBFL's root and model key}
\end{figure*}

\subsubsection{Key ratchets} 
The PQBFL integrates both symmetric and asymmetric ratchets to create secure communication in the FL protocol.
In each training round $r\in N$, the participants and server encrypt and decrypt the model information $Inf^{r}$ by using the model key $K_{i,j}$, where $i\in \ [1,L_{j}\ ]$ denotes the $i$-th symmetric ratchet within the $j$-th asymmetric ratchet.
The server determines the threshold for the symmetric ratchet range $L_{j}$ in each asymmetric ratchet based on its required efficiency and security.
The server can initiate a new period for the chain and model keys by deriving the root key $RK_{j+1}$ using the ratchet function $KDF_{A}$.
Subsequently, as illustrated in Figure \ref{fig:PQBFL's root and model key}, the updated or global model of each round is encrypted using distinct model keys $K_{i,j}$ derived from the previous chain key $CK_{i-1,j}$ using $KDF_{S}$.

The symmetric ratcheting counter $i$ increases until a new asymmetric ratcheting occurs, and after encryptions and decryptions in round $r$, the previous model key is discarded.
To perform asymmetric ratcheting, the server must regenerate and distribute new public KEM and ECDH keys to participants whose information payload $Inf_{b}^{r}$ in the publication step is updated to $<r,kpk_{b},epk_{b},M^{r-1},id_{p},id_{t},D_{t}>$.
The corresponding transaction \textit{Publish Task} is also converted to $Tx_{p}=\{r,h(Inf_{b}^{r}),h(kpk_{b}||epk_{b}),id_{p},id_{t},D_{t}\}$.
Upon receiving the new public keys, the participant derives a new root key, $RK_{j+1}$, and its subsequent model key, $K_{i,j+1}$, which is used for the next round encryption.
The information payload of the participant for transmission to the server is in the form $Inf_{a}^{r}: <r,ct,epk_{a},m^{r},id_{p},id_{t}>$, and the \textit{Update Model} transaction is $Tx_{u}=\{r,h(Inf_{a}^{r}),h(ct||epk_{a}),id_{p},id_{t}\}$.
These changes in the transaction and payload lead to the derivation of the same model key on both sides.

Here, we described how to establish keys and transmit models securely for a federated learning system using PQBFL. 
In the following section, we present a security analysis of our proposed protocol.

%============================
\section{Security analysis}
\label{Security Analysis}
%============================

%It is demonstrable that PQBFL satisfies the above-mentioned security standards.
In this section, we analyze the security of the proposed protocol given the assumptions of cryptographic hardness problems. 
We present a security model to analyze the security of the PQBFL. 
It relies on a multistage Authenticated Key Exchange (AKE) security model called Fischlin and Göunther \citep{Multi-stage_key_exchange}.
The basic concept is that an adversary should communicate with a challenger who represents both parties. 

\subsection{Security Model} 
In this security model, we address the confidentiality, forward secrecy, and post-compromise security of the proposed scheme against quantum powered attackers. 
The adversary models a passive or active opponent by directing every interaction between the truthful parties. 
Furthermore, the adversary can obtain root keys and ephemeral shared secrets to create model keys among other secrets belonging to truthful parties. 
Next, an honest party challenges the adversary in obtaining information regarding the model key.  This is modeled as the adversary's ability to differentiate the model key from a completely random bit string of the same length.  
If the adversary cannot do so with a probability significantly different from $\frac{1}{2}$, then the model keys are random for the adversary.

Let $\Pi$ be a post-quantum security protocol that uses KEM for confidentiality. 
The notion of security is defined in terms of the adversary’s advantage in a security experiment. 
The advantage of adversary $\mathcal{A}$ in the security experiment is defined as:

\DeclarePairedDelimiter{\abs}{\lvert}{\rvert}
\begin{equation}
    Adv_{\Pi,\mathcal{A}}^{ind-cca}(\lambda)=\abs{Pr[b'=b]-\frac{1}{2}}
\end{equation}

\noindent where the probability is assumed by the randomness of the challenger $\mathcal{C}$ and the adversary $\mathcal{A}$.

Theorem 1 demonstrates that the security of the model keys in the PQBFL is secure in a hybrid traditional and post-quantum environment. 
 This means that they are secure if either the ECDH problem assumptions remain difficult or the post-quantum scheme remains secure. This can deduced from the fact that the pertinent advantage boundaries contain a term of the type 
 
$$
    \min \{ \epsilon_{G}^{\text{PRF-DH}} + \ldots, + \epsilon_{\Pi}^{\text{IND-CCA}} + \ldots \}
$$

This indicates that security remains valid even if one of the two foundational assumptions is compromised, provided that the other assumptions are unbroken.

\textbf{Theorem 1}:
Assume that $\textit{Expand}, \textit{Extract}$ algorithms that constitute $\text{KDF}_{S}$ and $\text{KDF}_{A}$ ratcheting processes are all PRFs in their arguments and $\sigma$ is an EUF-CMA secure signature.  
We also assume that $\Pi$ is an IND-CCA-secure KEM, and the PRF-ODH-secure assumption applies to the elliptic curve group.    
The PQBFL considers that the $i$-th model key in the symmetric ratchet that arises from the $j$-th asymmetric ratchet in round $r$ is fresh. Consequently, the derived model key $K_{i,j}$ is indistinguishable from a random bit string of identical length.

\textbf{Theorem 2}:
In the PQBFL protocol, if the cryptographic primitives (KDF, IND-CCA secure KEM, and PRF-ODH secure ECDH) are secure, then any session key $K_{i,j}$ derived within the symmetric ratcheting mechanism is computationally independent of prior keys (forward secrecy), and future keys remain indistinguishable from random even after compromising the current key when asymmetric ratcheting occurs (post-compromise security).

\subsection{Security Proof} 

\noindent\textbf{Confidentionlity.} 
We prove the security of the root key $RK_{j}$ and the chain key $CK_{i,j-1}$ used to derive a particular model key $K_{i,j}$.
Furthermore, because KEM and ECDH techniques are primarily responsible for the initial establishment of keys, we must analyze and validate long-term security using theorems, lemmas, and games. 
The following are the lemmas for each phase:

\begin{itemize}
    \item \textbf{Lemma 1} indicates that the root key $RK_{0}$ and the chain key $CK_{0.0}$ established by the server and participant sides are secure during the initial establishment of the key. 
    \item \textbf{Lemma 2} demonstrates the security of the chain keys $CK_{0,j}$ and the root $RK_{j}$ established by the asymmetric ratchet on the participants and server sides.
    \item \textbf{Lemma 3}. shows that the chain keys $CK_{i,j}$ and the model keys $K_{i,j}$ derived in the symmetric ratchet are secure.
\end{itemize}

%Proof: Assume that an adversary $\mathcal{A}$ can distinguish the root key and chain key from a uniformly random bit-string of the same length with a non-negligible advantage $\epsilon$. We will construct an algorithm $\mathcal{B}$  that can break the security of the KEM or the KDF with the same advantage $\epsilon$ which contradicts the assumption that the KEM and KDF are secure.$\mathcal{B}$ receives public key $epk$ and $kpk$ from the challenger and sends to $\mathcal{A}$ as the server's public key.

\textbf{Proof phase:} we provide a sequence of games $G$ that gradually modify the security experiment to simplify the analysis. 

%We show that advantage of $\mathcal{A}$ in winning the IND-CCA security game is negligible, where G0 is the original IND-CCA game. The adversary $\mathcal{A}$ chooses two plaintext messages $m_{0}$ and $m_{1}$ of the same length and sends them to the challenger $\mathcal{C}$. The challenger $\mathcal{C}$ chooses a bit $b \leftarrow{0,1}$ at random, encapsulates $m_{b}$ under the public key $pk$ of an honest party to obtain the encapsulation $c^{*}$ and the corresponding decapsulation key $sk$, and sends $c^{*}$ to the adversary.Let $q_{e}$, $q_{d}$, $q_{k}$ and $q_{h}$ be encapsulation, decapsulation, key derivation and hash queries that the adversary $\mathcal{A}$ can make, respectively.

\textbf{G0:} This is the original IND-CCA security game between the adversary $\mathcal{A}$ and the challenger $\mathcal{C}$ in PQBFL. 

$$
    Adv_{0}= Adv_{\Pi}^{ind}(\mathcal{A})
$$

\textbf{G1:} In this game, the challenger $\mathcal{C}$ substitutes a random bit string of the same length for the $SS_{k}$ value. 
The adversary $\mathcal{A}$ receives as input a challenge cipher-text $ct^{*}$ where $ct\neq ct^{*}$, random-or-real value $SS_{k}^{*}$, and decapsulation oracle $O(Decap)$  using the private key $ksk$, and then must return its estimate of whether $SS_{k}^{*}$ is random or real. 
The adversary $\mathcal{A}$ outputs as its response to the IND-CCA challenger the same $b' \in \{0,1\}$ output. In the case of a real $SS_{k}^{*}$, $b'=0$, whereas when $SS_{k}^{*}$ is random, $b'=1$. 
Thus

$$
    Adv_{0} \leq Adv_{\Pi}^{ind}(\mathcal{A})+ Adv_{1}
$$

%The challenger generates a key pair $(pk,sk)$, and sends the $pk$ to the $\mathcal{A}$. The adversary can make encapsulation queries to the challenger, and can also request the challenge cipher-text. When the adversary requests the challenge cipher-text the challenger chooses a random bit $b \in \{0,1\}$ generates a shared secret $(ss,ct) \leftarrow Encap(pk)$ and sends $ct$ to the $\mathcal{A}$. The adversary's goal is to guess the value of $b$. Let $Adv_{\Pi,\mathcal{A}}^{G0}(\lambda)$ denote the advantage of the adversary in Game 0, where $\lambda$ is the security parameter.

\textbf{G2:}  In this game, the challenger $\mathcal{C}$ generates the initial root keys $RK_{0}$ and chain keys $CK_{0.0}$ for the several rounds. 
The challenger replaces the output of $\text{KDF}_{A}$ the used in the asymmetric ratchet phase with a random bit string with the same length $RK^{*}_{0}$.
Adversary $\mathcal{A}$ receives the real or random output of $\text{KDF}_{A}$ and interacts with the system to distinguish whether the output is real or random. 
The adversary outputs $b'\in \{0,1\}$ as a guess of whether the output is real or random. 
When $b'$ output is real, it simulates Game 1 since $SS$ is random, whereas when output is random, it simulates Game 2 so that adversary's advantage in distinguishing the $KDF_{A}$ output in game is

$$
    Adv_{1} \leq Adv_{prf}^{kdf_{A}}(\mathcal{A})+ Adv_{2}
$$

%the challenger uses a random shared secret $ss^{*}$ instead of the actual shared secret $ss$ in the challenge phase. When the adversary $\mathcal{A}$ requests the challenge cipher-text $ct$ which $ct\neq ct^{*}$, the challenger chooses a random bit $b \in \{0,1\}$, generates a random shared secret $(ct^{*},ss^{*}) \leftarrow Encap(pk)$ and send $ct^{*}$ to $\mathcal{A}$. The adversary's view in this game is identical to the view in G0, since the distribution of $ss^{*}$ is identical to distribution of $ss$.

\textbf{G3:} 
Challenger $\mathcal{C}$ generates initial keys and states for the symmetric ratchet. 
Subsequently, the challenger replaces the output of $\text{KDF}_{S}$ (chain keys $CK_{i,j}$ and model key $K_{ij}$) with random bit strings. 
For each derivation of $CK_{i,j}$ and $K_{i,j}$, the adversary $\mathcal{A}$  interacts with the protocol as usual and tries to distinguish the real keys from random ones with output guess $b' \in \{0,1\}$. 
Thus, the adversary's advantage $G2$ is bounded by the PRF security of $\text{KDF}$. 
By combining these, we can conclude that the adversary's advantage in distinguishing the real chain keys $CK_{i,j}$ and model key $K_{i,j}$ from random bit strings is negligible.

$$
    Adv_{2} \leq Adv_{CK_{ij}}^{kdf_{s}}(\mathcal{A})+ Adv_{K_{ij}}^{kdf_{s}}(\mathcal{A})+ Adv_{3}
$$

Therefore, the chain keys $CK_{i,j}$ and the model key $K_{ij}$ derived in the symmetric ratchet are secure. The above games show that the PQBFL achieves confidentiality under adaptive chosen cipher-text security.

\noindent\textbf{Forward and Post-compromise security.}
In PQBFL, the server and participants utilize a fresh model key $K_{i,j}$ for each training round derived through KDF symmetric ratcheting.  
Forward secrecy ensures that if the $r$-th round key is compromised, keys from earlier rounds ($K_{i\ ',j}$ for $i\ '<i$) are secure, thereby securing the previously transmitted models. 
The symmetric ratcheting mechanism ensures that each key $K_{i,j}$ derived from a fresh chain key $CK_{i-1,j}$ is secure and is discarded after use.  

$$     
    CK_{i,j}=KDF_{s}(CK_{i-1,j},info) 
$$  
 
The adversary’s advantage in compromising earlier keys, given $K_{i,j}$ is  
 
$$     
    Adv_{\mathcal{A}}^{forward} \leq Adv_{\mathcal{A}}^{PRF} 
$$ 

In addition, the server can trigger asymmetric ratcheting by sending new public keys to participants to derive a new root key $RK_{j+1}$.   
Whenever a new root key is established, it provides post-compromise security for the model transmission. 
Post-compromise security ensures that if a key $K_{i,j}$ is compromised, the future keys remain secure. 
Asymmetric ratcheting triggers a new key exchange when the threshold $L_{j}$ of the symmetric ratcheting range is reached:

$$      
    RK_{j+1}=KDF_{A}(SS'_{k},SS'_{e}) 
$$  
where  $SS'_{k}$ and $SS'_{e}$ are derived from freshly generated public-private key pairs. 
The adversary’s advantage in compromising the future keys given $K_{i,j}$ is  

$$     
    Adv_{\mathcal{A}}^{post-compromise} \leq min\{ Adv_{\mathcal{A},Kyber}^{IND-CCA}, Adv_{\mathcal{A}}^{PRF-DH}\} 
$$ 

In other words, given that $K_{i,j}$ represents the model key for the $i$-th symmetric and $j$-th asymmetric ratchet, if key $K_{i,j}$ is compromised, the set of compromised keys $K_{c}$ can be defined as  
$$     
    K_{c}=\{K_{i+n,j} \ | \ 0\leq n \leq L_{j}\} 
$$  
\noindent where $n$ denotes the number of compromised model keys. This indicates that, if the model key $K_{i,j}$ is compromised, the attacker can only retrieve the models encrypted by the model keys up to $K_{i,j+1}$ with the following probability:  

$$ 
    P[Compromise(K_{i+n,j})\ | K_{c}] 
$$

In the following section, we discuss the security attacks and threats mitigated by PQBFL.

\subsection{Security discussion}

In this section, we explain how PQBFL meets the security requirements mentioned in Section \ref{Security and Privacy requirements}. 
These requirements are covered differently in this context. For instance, dual-layer security and resistance against various attacks can be used to examine confidentiality.

\noindent\textbf{Authentication.}
For the authenticity of the received payloads in the off-chain channel, the PQBFL employs identical ECDSA signature keys used for registration transactions. 
The participants and server first made a registration transaction using their private keys on the blockchain. 
Given the immutability of registered transactions, using different public keys used in registration for signing during data transmission makes it impossible to verify the sender. 
This approach ensures that adversaries cannot sign data with a different public key and guarantees the integrity of authentication.

\noindent\textbf{Dual-Layer security.}
Although post-quantum schemes are analyzed and standardized by the cryptography community, they are in their infancy and we cannot trust them completely. 
PQBFL adopts a hybrid approach using both ECDH and Kyber to derive root keys; consequently, the attacker must break both classical and post-quantum algorithms. 
This makes PQBFL secure against HNDL attacks and mitigates potential vulnerabilities in post-quantum standard algorithms.

\noindent\textbf{Identity privacy vs traceability.}
It should be noted that a reputation-based mechanism requires traceability of rewards and penalties. This makes it impossible to simultaneously achieve complete anonymity and traceability in such systems. 
However, PQBFL utilizes blockchains that offer pseudo-anonymity via blockchain addresses, thereby providing the identity privacy and traceability of participants in an FL system. 

\noindent\textbf{Resistance of Various Attacks.}
We found that PQBFL provides security against various attacks in FL environments.
\begin{enumerate}
    \item \textit{Free-riding attack:} The strong encryption and transaction records of both local and global models between the server and participants indicate that no entity can gain unauthorized access to the models, thus preventing a free-riding attack.
    \item \textit{DDoS attack:} The registration phase and the recording of different transactions on the distributed blockchain within the PQBFL eliminate the possibility of a DDoS attack.
    \item \textit{Man-in-the-middle attack:} The system's ability to provide secure authentication is evident from the previous explanation. 
    Consequently, this mitigates the impact of man-in-the-middle attacks.
    \item \textit{Replay attack:} Because the participants and the server generate a fresh model key and use a unique timestamp for each training round, the members of an FL project can easily detect any replay attack.
\end{enumerate}

The features discussed here demonstrate that the proposed scheme can withstand several attacks and security risks in FL environments. 
In the next section, we examine the performance of the PQBFL.

%============================
\section{Performance Analysis}
\label{Performance Analysis}
%============================
To provide a balance between efficacy and security, PQBFL presents forward secrecy and post-compromise security with a ratcheting mechanism, whereas other studies must exchange keys for each round to provide this capability, which has an extremely high computational overhead.
More details on the implementation of the protocol, including server and client source codes, are available on GitHub \citep{source_code}. 
In the following section, we analyze computational and communication costs and evaluate the proposed scheme.

\subsection{Computation Cost}
The computation cost of each party (server or participant) can be broken down as follows: 1) key establishment between server and participant, including key generation, encapsulation, and decapsulation operations, which is $O(n)$ and 2) key derivation, encryption, signing, and verification, $O(n)$. 
Overall, the computational complexity for each party is $O(n^2)$ for all rounds in an FL project. 
However, the total costs for the participants and the servers are different, as shown in Table \ref{tab: Operation time} because only the encapsulation operation is on the participant side, and the Kyber key generation and decapsulation are located on the server side. 
This shows that the participants require fewer computational operations, which is desirable for devices with constrained resources.

\begin{comment}
    \begin{figure*}[h!]
    \centering
    \subfloat[Participant]{\includegraphics[width=2.2in,height=1.9in]{Images/computation_participant3.png}}%
    \hspace{0.001\linewidth} % Add small horizontal spacing
    \subfloat[Server]{\includegraphics[width=2.2in,height=1.9in]{Images/computation_server1.png}}%
    \hspace{0.001\linewidth} % Add small horizontal spacing
    \subfloat[Total]{\includegraphics[width=2.2in,height=1.9in]{Images/computation_total.png}}%
    \caption{Computation performance in different symmetric ratcheting ranges}
    \label{fig: symmtric_ratcheting_range}
    \end{figure*}
\end{comment}

\begin{figure*}[h!]
    \centering
    \begin{subfigure}[b]{0.3\textwidth}
        \centering
        \includegraphics[width=2.2in,height=1.9in]{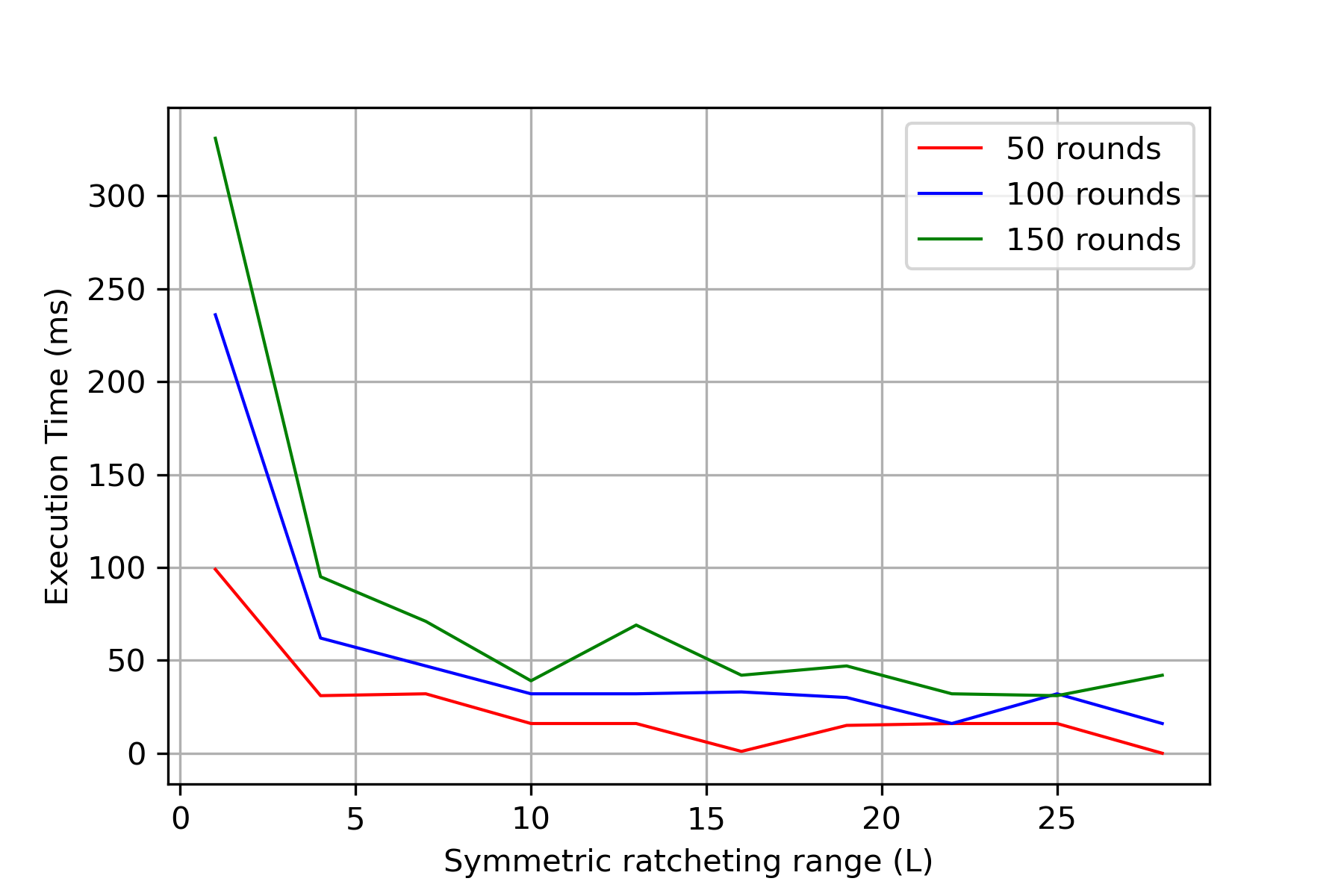}
        \caption{Participant}
    \end{subfigure}
    \hspace{0.001\linewidth}
    \begin{subfigure}[b]{0.3\textwidth}
        \centering
        \includegraphics[width=2.2in,height=1.9in]{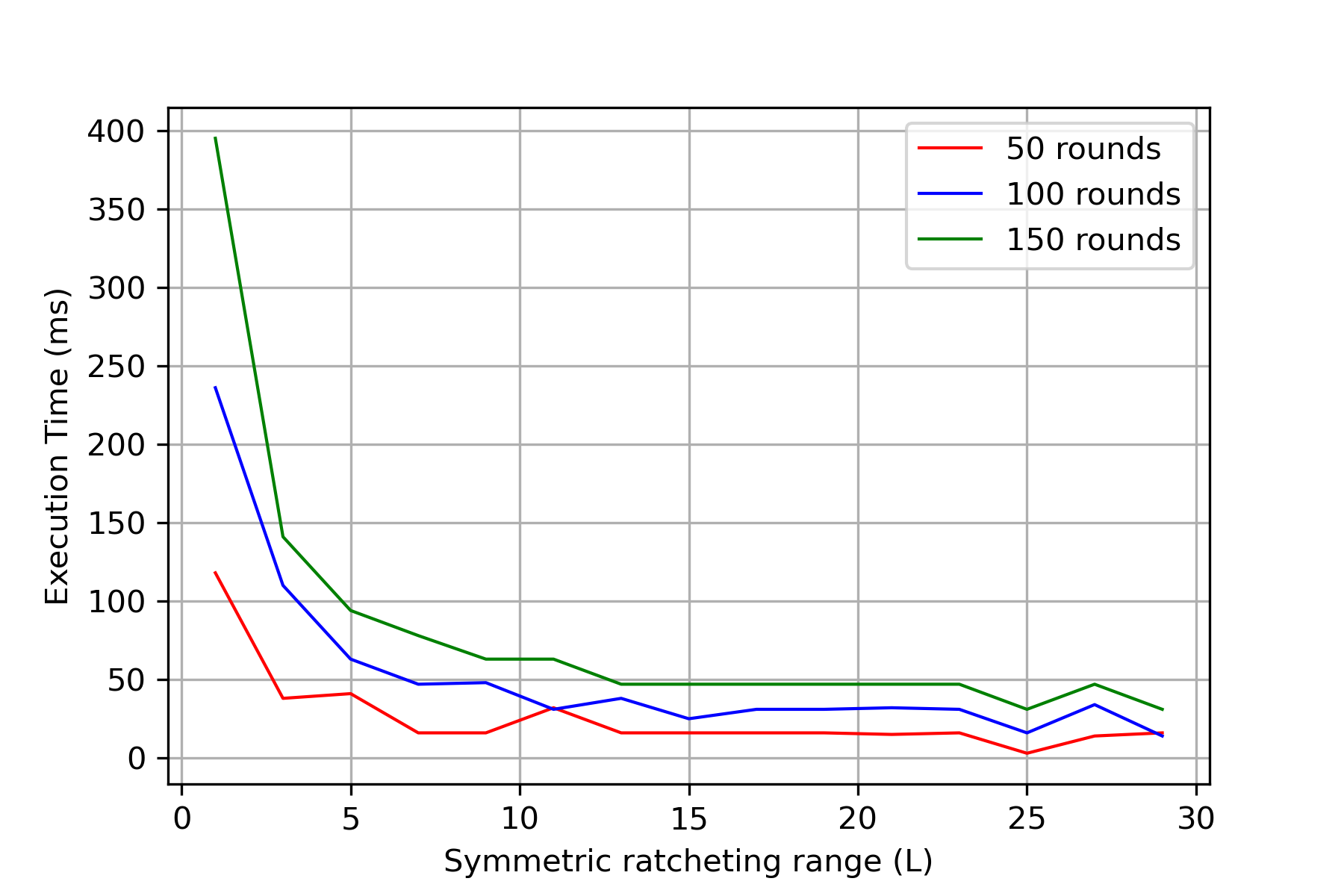}
        \caption{Server}
    \end{subfigure}
    \hspace{0.001\linewidth}
    \begin{subfigure}[b]{0.3\textwidth}
        \centering
        \includegraphics[width=2.2in,height=1.9in]{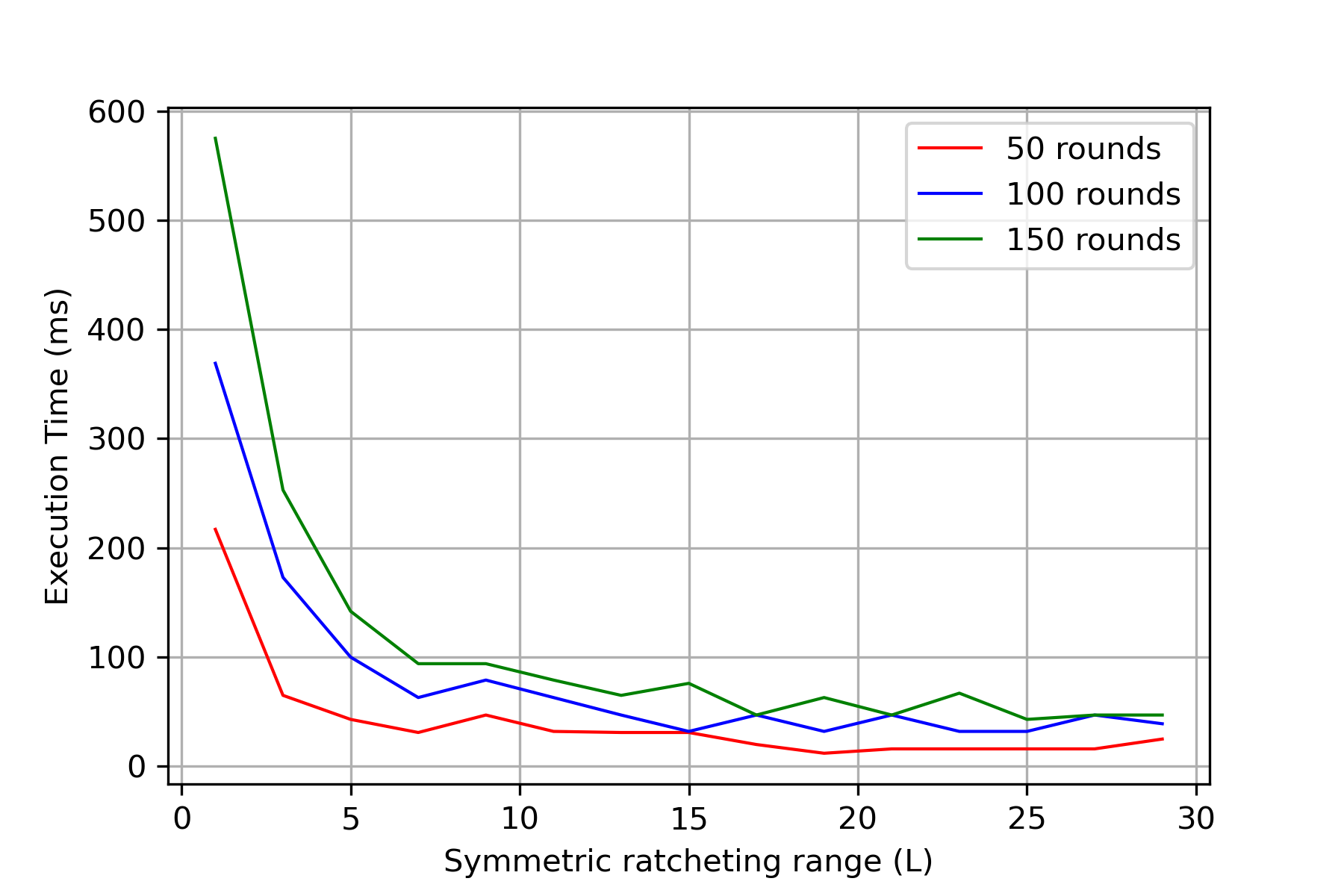}
        \caption{Total}
    \end{subfigure}
    \caption{Computation performance in different symmetric ratcheting ranges}
    \label{fig: symetric_ratcheting_range}
\end{figure*}

We evaluated the performance using Post-Quantum Cryptography (PQCrypto) \citep{pqcrypto_library} and PyCryptodome \citep{pycryptodome} Python libraries, and the execution time of cryptographic operations was measured on running Windows 11 with an Intel(R) Core(TM) i7-1195G7 @ 2.90GHz and 32GB of RAM.
To provide a balance between security and efficiency, we considered a 128-bit quantum security level and selected the following cryptographic parameters:

\begin{itemize}
    \item We utilize  kyber-768 for KEM and ECDH over the NIST P-256 curve for key exchange,
    \item Key derivation functions instantiate with HKDF using SHA-384 hash functions.
    \item We use ECDSA over secp256k1 as blockchain and models signature algorithm and AES-256 for model encryptions.
\end{itemize}

\begin{table}[ht]
\centering
\caption{Comparison of first two rounds operation time in both side}
\label{tab: Operation time}
\begin{adjustbox}{width=3.2in}
\begin{tblr}{
  row{1} = {c},
  row{2} = {c},
  cell{1}{1} = {r=2}{},
  cell{1}{2} = {c=2}{},
  cell{1}{4} = {r=2}{},
  cell{3}{2} = {c},
  cell{3}{3} = {c},
  cell{3}{4} = {c},
  cell{4}{2} = {c},
  cell{4}{3} = {c},
  cell{4}{4} = {c},
  hline{1,3,5} = {-}{},
  hline{2} = {2-3}{},
  %hline{3} = {2-4}{},
}
Party       & Operations                 &                & {Total \\cost (ns)} \\
            & Asymmetric ratchet rounds               & Symmetric ratchet rounds  &                    \\
Participant & $T_{kg}+T_{enc}+2T_{dr}+T_{s}+T_{v}$  & $2T_{dr}+T_{s}+T_{v}$ & 4408                \\
Server      & $2T_{kg}+T_{dnc}+2T_{dr}+T_{s}+T_{v}$ & $2T_{dr}+T_{s}+T_{v}$ & 5065               
\end{tblr}
\end{adjustbox}
\end{table}

Public-key cryptography algorithms are used primarily to establish a new root key $RK_{j}$ through asymmetric ratcheting. 
The server determines the root-key lifetime based on the required security level by selecting an appropriate symmetric ratcheting range $L_{j}$. 
To evaluate the impact of different symmetric ratcheting ranges on computational costs over varying rounds, we performed experiments using various values $L_{j}$, as illustrated in Figure \ref{fig: symetric_ratcheting_range}. 
The analysis was performed for the participant, the server and the total computation costs of both parties combined.
The results demonstrate that as the symmetric ratcheting range expands, the execution time required to compute the model key decreases. This reduction is attributed to the decreased reliance on public-key operations. 
In particular, the server generally incurs slightly higher computational costs compared to the participant. 
This difference arises from the server's need to perform both post-quantum key generation and decapsulation, while the participant only handles encapsulation. This distinction is evident in the graphs.
In addition, in some cases, the computation time curves for 50, 100, and 150 rounds overlap. 
This behavior is due to the inherent use of pseudo-random functions in the generation of post-quantum and traditional keys, leading to similar computational patterns across these settings. 
These findings suggest that an ideal symmetric ratcheting range to balance security and computational efficiency is approximately 10 symmetric ratchets per asymmetric ratchet.

\subsection{Communication Cost}
The aforementioned studies assumed only a few training rounds for the FL systems. 
However, practical FL training requires several rounds to achieve complete convergence in the global model. 
Thus, they need to exchange keys in every round for forward and post-compromise security, which creates communication and computational overhead.  
We used a hybrid procedure and do not exchange public keys or models through the blockchain network for key establishment because it is not cost-effective. 
We broadcast the hash of the information keys and model, which can significantly decrease the communication costs of the blockchain. 
However, the communication costs for key exchanges between participants and servers are divided into two categories: on-chain and off-chain costs.  
A participant receives two public keys from the server and sends its public key and ciphertext to the server via the off-chain channel at $O(n)$ cost.
Moreover, the on-chain channel includes sending and receiving \textit{Registration} transactions with $O(m)$ communication costs. 
Therefore, the overall communication cost of PQBFL for each party is $O(n+m)$.

\begin{figure}[ht]
    \centering
\includegraphics[width=3.2in,height=2.3in]{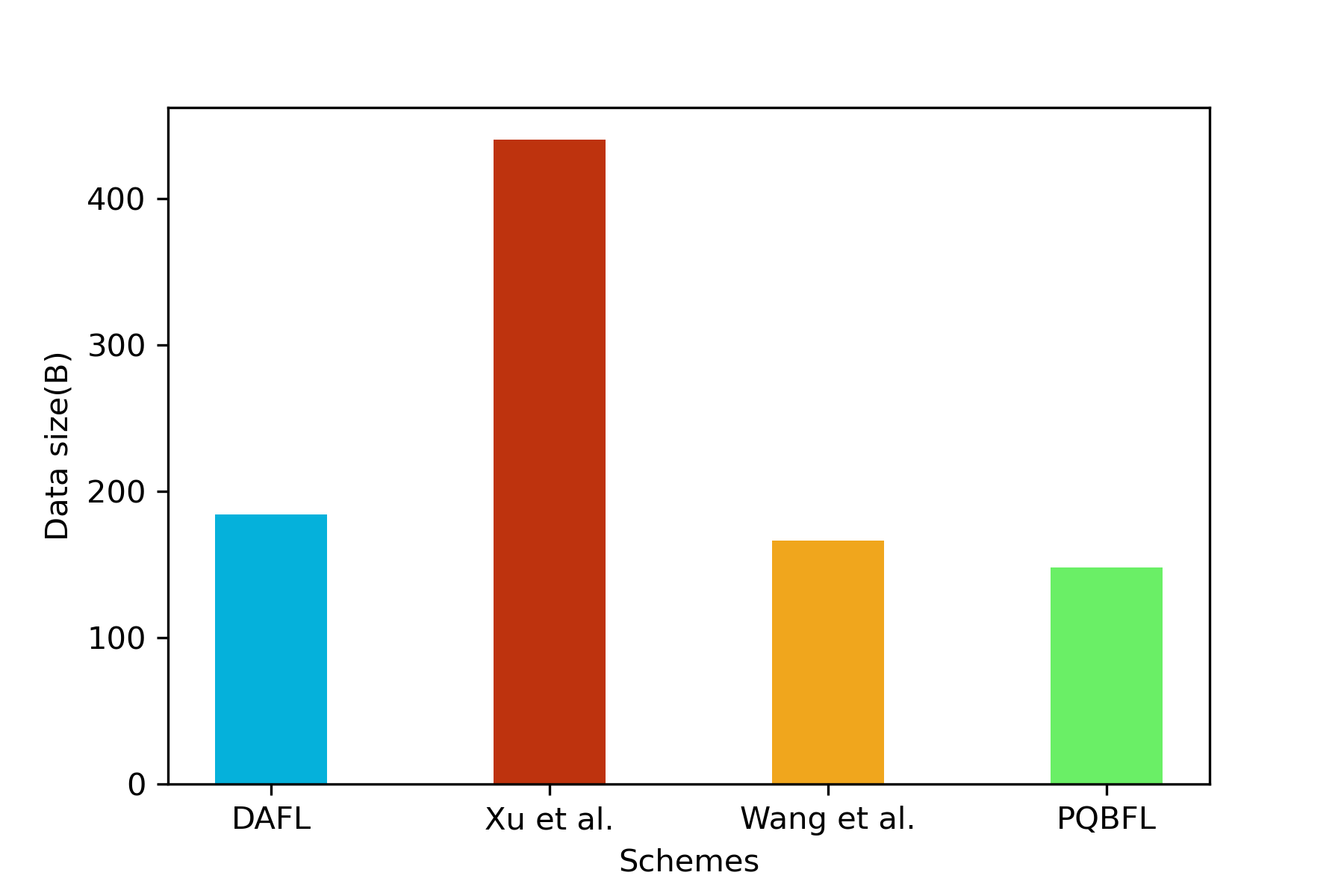}
    \caption{Comparison of communication costs on blockchain.}
    \label{fig: compare data on-chain}
\end{figure}

In our scheme, \textit{Registration} transactions that occur once during an FL project require only $3(32)+2(2)=100$ byte data for both the server and a single participant. 
In addition, the size of the \textit{Publish, Update} and \textit{Feedback} transactions in the whole of a single round is only $4(32)+3(4)+2(2)+4(1)=148$ bytes, while, as can be seen in Figure \ref{fig: compare data on-chain} DAFL\citep{Blockchain-based_decentralized}, \citep{A_certificateless_signcryption} and \citep{Secure-enhanced} require 184, 440, and 166 bytes only for the update message on the blockchain. 
These 148 bytes include 32 bytes for the hash of the models and keys and four bytes for identifiers in all three transactions. 
Two bytes are used for the deadline task and score in \textit{Publish, Feedback}, one byte for the round numbers in the three transactions, and a termination flag in the \textit{Feedback} transaction.

The gas consumption for various transactions in a single round of the PQBFL protocol is presented in Table \ref{tab: Gas cunsumption}. 
The analysis highlights that the participant incurs a total gas cost of 305.724 units, mainly attributed to the registration (72.452 units) and update model (233.272 units) transactions. 
Meanwhile, the server has a higher total gas consumption of 515.932 units, distributed through registration (236.533 units), publish task (258.329 units), and feedback model (211.070 units). These results reflect the computational overhead for both parties, with the server bearing a greater burden due to its role in task publication and managing feedback for model updates.

\begin{comment}
\begin{figure*}[b]
    \centering
    \subfloat[$L_{j}=5$]{\includegraphics[width=2.2in,height=1.9in]{Images/communication_cost5.png}}%
    \hspace{0.001\linewidth} % Add small horizontal spacing
    \subfloat[$L_{j}=10$]{\includegraphics[width=2.2in,height=1.9in]{Images/communication_cost10.png}}%
    \hspace{0.001\linewidth} % Add small horizontal spacing
    \subfloat[$L_{j}=20$]{\includegraphics[width=2.2in,height=1.9in]{Images/communication_cost20.png}}%
    \caption{Size of off-chain data transmitted in three different symmetric ratcheting ranges.}
    \label{fig: size of data transmitted}
\end{figure*}
\end{comment}

\begin{figure*}[b]
    \centering
    \begin{subfigure}[b]{0.3\textwidth}
        \centering
        \includegraphics[width=2.2in,height=1.9in]{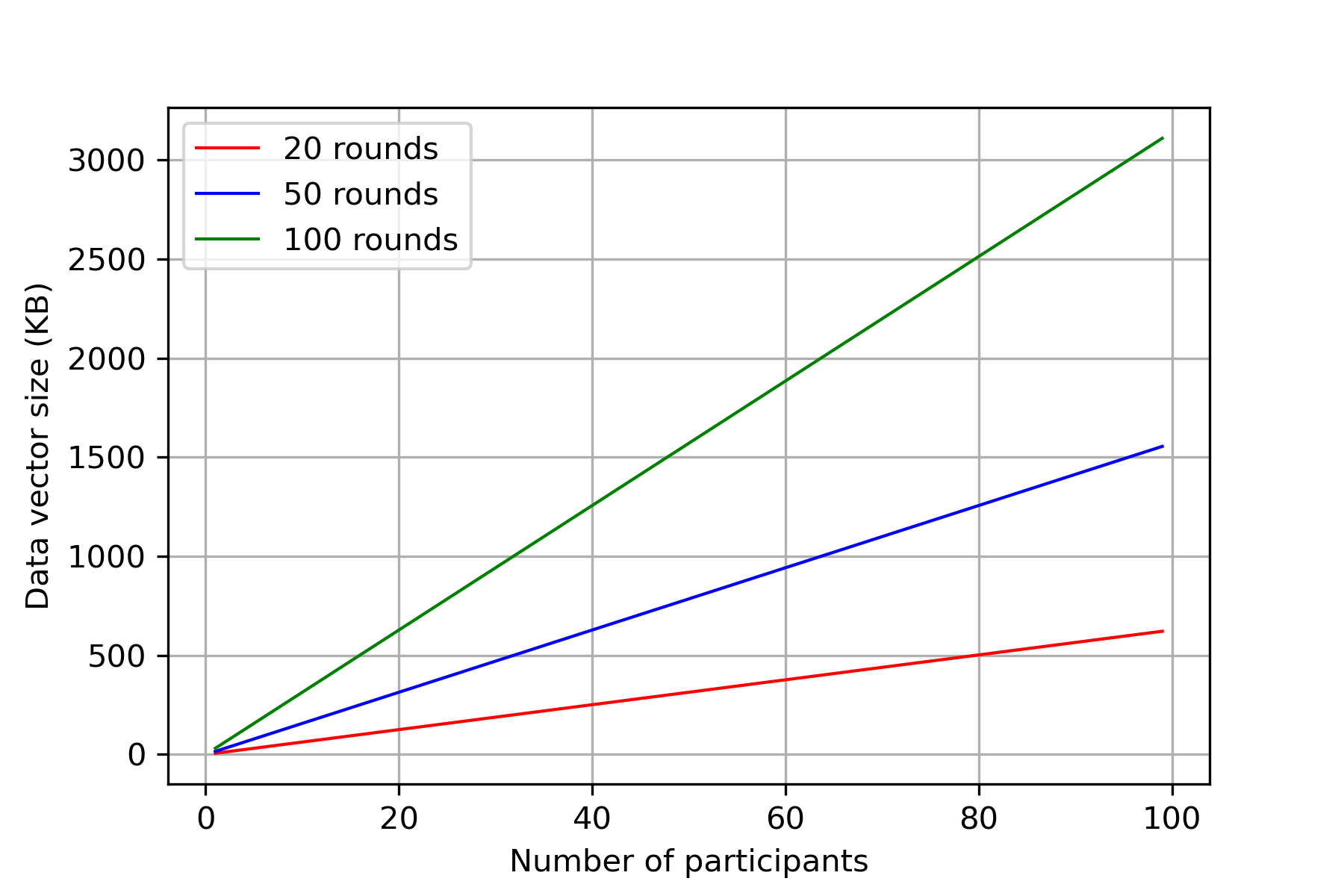}
        \caption{$L_{j}=5$}
    \end{subfigure}
    \hspace{0.001\linewidth}
    \begin{subfigure}[b]{0.3\textwidth}
        \centering
        \includegraphics[width=2.2in,height=1.9in]{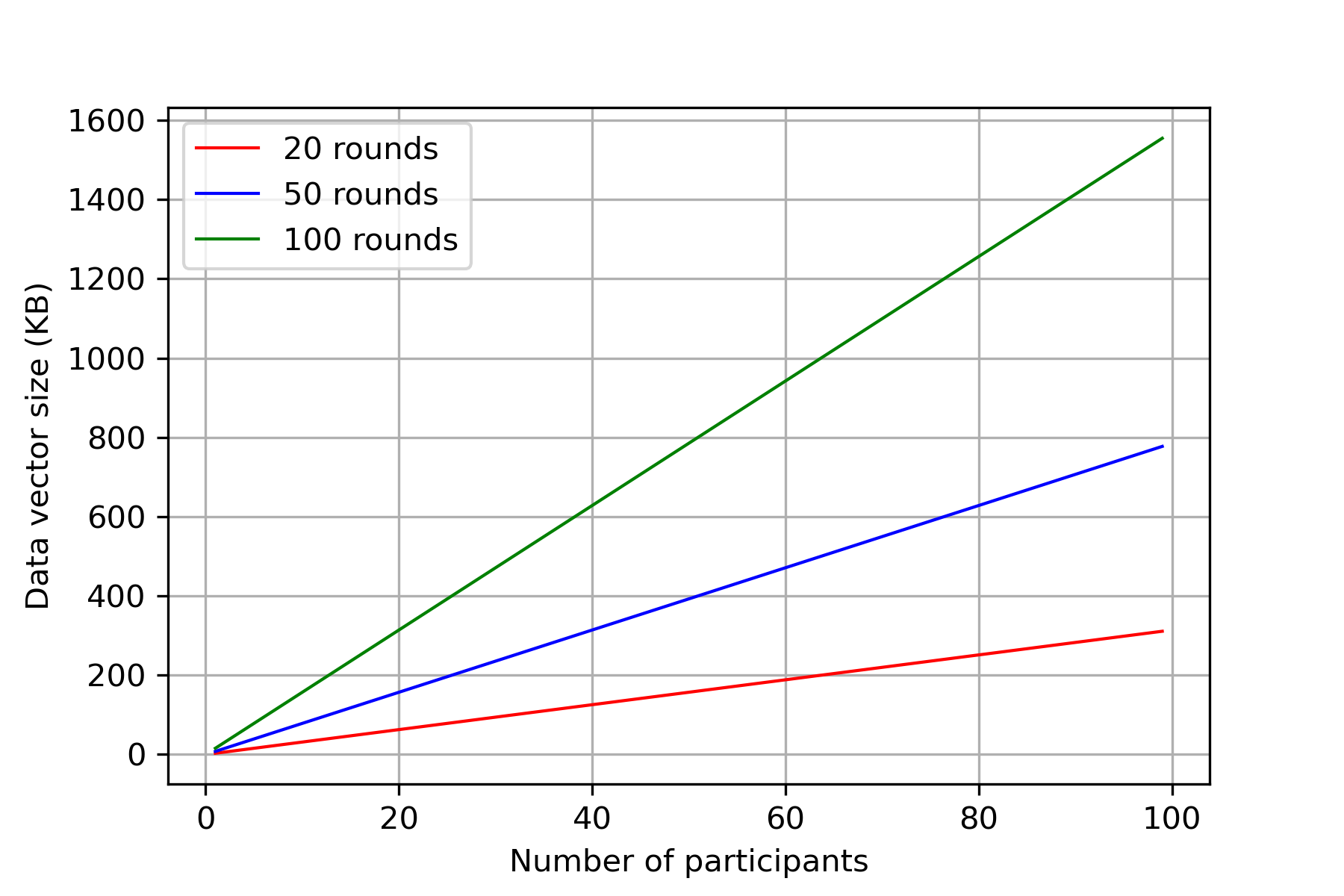}
        \caption{$L_{j}=10$}
    \end{subfigure}
    \hspace{0.001\linewidth}
    \begin{subfigure}[b]{0.3\textwidth}
        \centering
        \includegraphics[width=2.2in,height=1.9in]{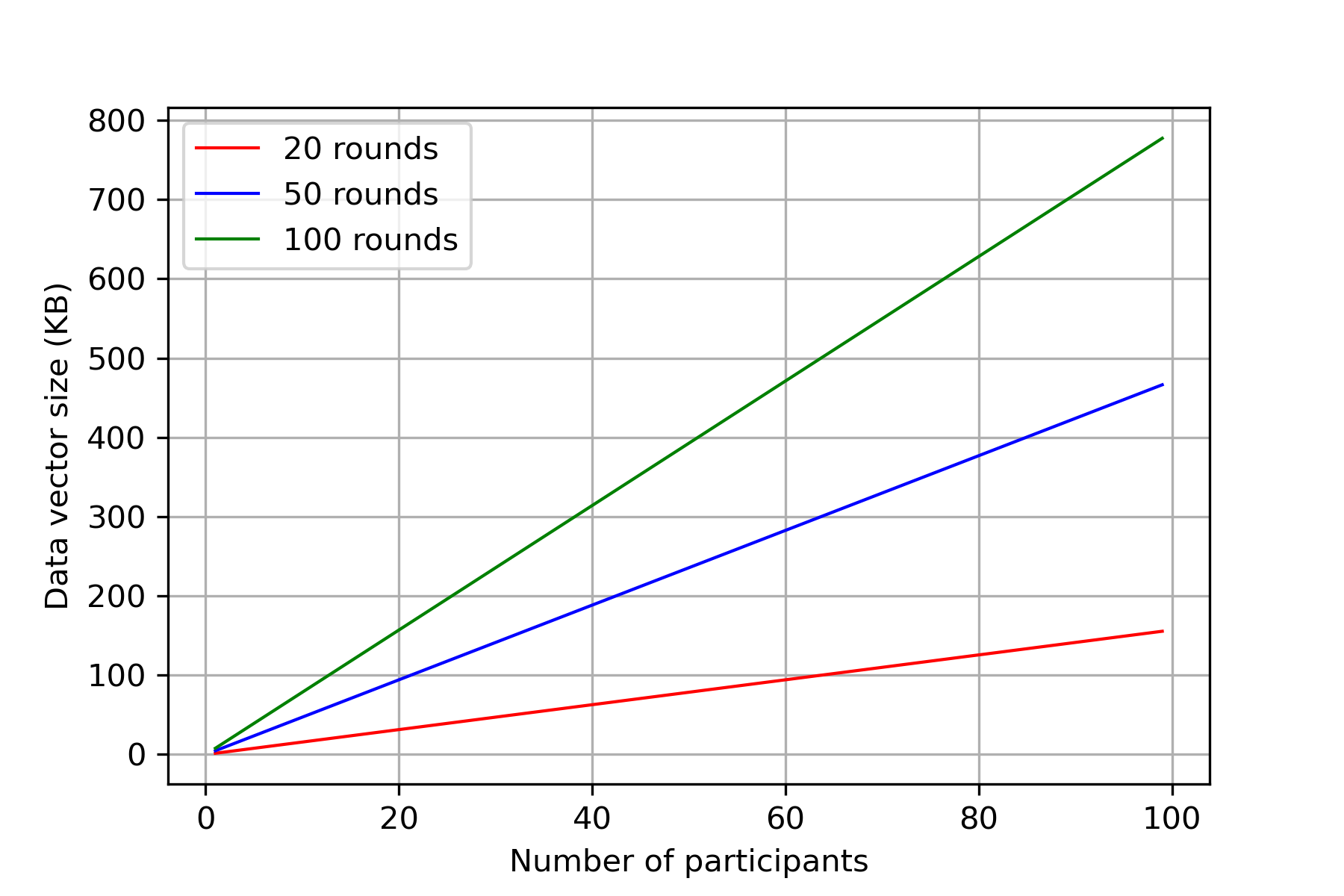}
        \caption{$L_{j}=20$}
    \end{subfigure}
    \caption{Size of off-chain data transmitted in three different symmetric ratcheting ranges.}
    \label{fig: size_of_data_transmitted}
\end{figure*}

\begin{table}
\centering
\caption{Gas consumption in different transactions in a single round}
\label{tab: Gas cunsumption}
\begin{adjustbox}{width=3.2in}
\begin{tblr}{
  row{1} = {c},
  row{2} = {c},
  cell{1}{1} = {r=2}{},
  cell{1}{2} = {c=4}{},
  cell{1}{6} = {r=2}{},
  cell{3}{2} = {c},
  cell{3}{3} = {c},
  cell{3}{4} = {c},
  cell{3}{5} = {c},
  cell{3}{6} = {c},
  cell{4}{2} = {c},
  cell{4}{3} = {c},
  cell{4}{4} = {c},
  cell{4}{5} = {c},
  cell{4}{6} = {c},
  hline{1,3,5} = {-}{},
  hline{2} = {2-5}{},
}
Party       & Transactions &              &              &                & {Total \\Gas} \\
            & Registration & Publish Task & Update model & Feedback model &               \\
Participant & 72.452       & -            & 233.272      & -              & 305.724       \\
Server      & 236.533      & 258.329      & -            & 211.070        & 515.932       
\end{tblr}
\end{adjustbox}
\end{table}

We also evaluated the communication performance of the PQBFL in the presence of various participants, as shown in Figure \ref{fig: size_of_data_transmitted}. 
As illustrated, the fixed communication cost on the off-chain channel increases linearly with the number of participants across different rounds. However, using a smaller symmetric ratcheting range significantly increases the amount of data transmitted due to the more frequent public key exchanges. 
Notably, as the symmetric ratcheting range expands to $L_j=20$, the transmitted data is reduced to less than one-third of the amount required for a smaller range, such as $L_j=5$.

\begin{figure}[ht]
    \centering
    \begin{subfigure}[b]{0.23\textwidth}
        \centering
        \includegraphics[width=1.6in,height=1.6in]{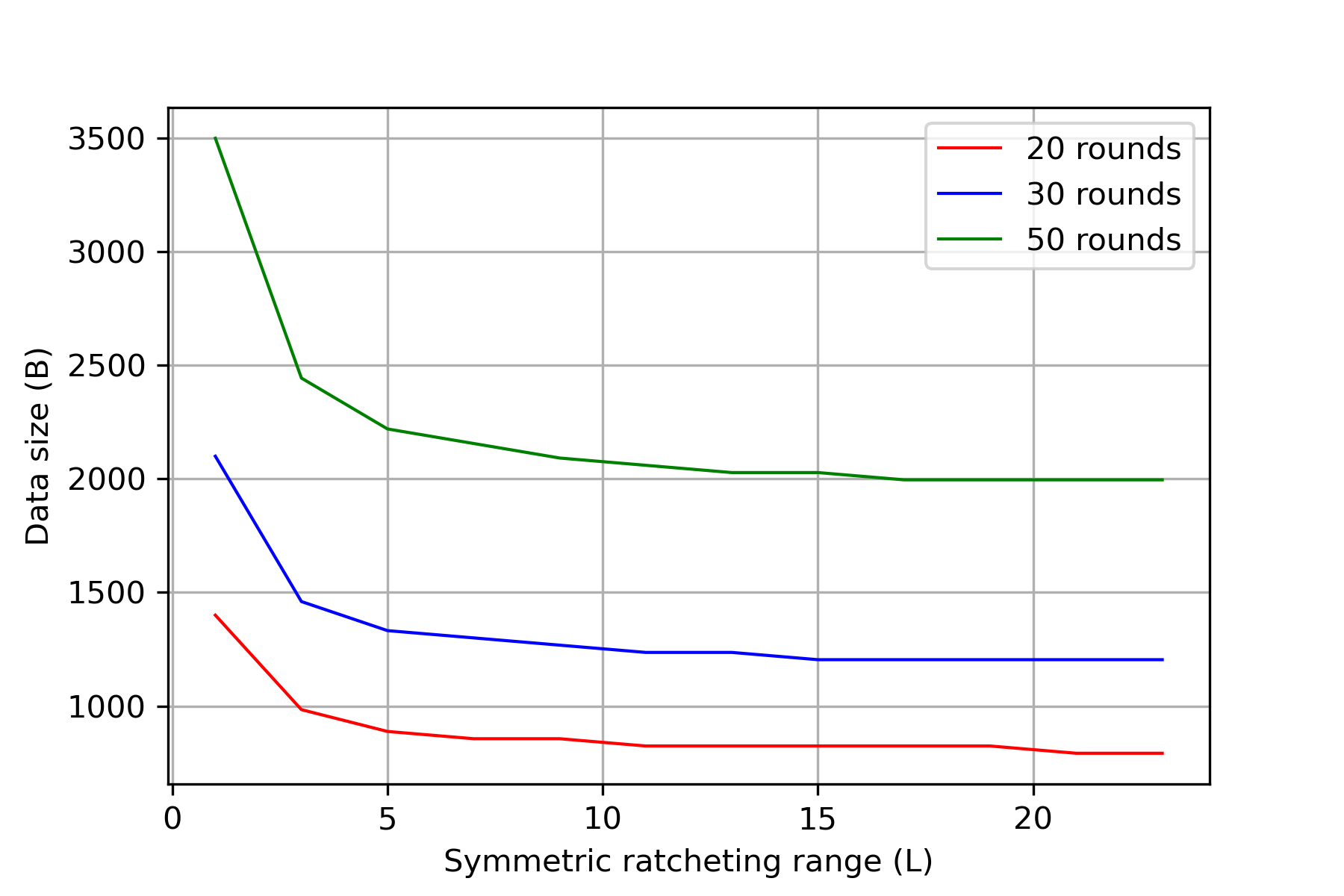}
        \caption{Participant}
    \end{subfigure}
    \hspace{0.001\linewidth}
    \begin{subfigure}[b]{0.23\textwidth}
        \centering
        \includegraphics[width=1.6in,height=1.6in]{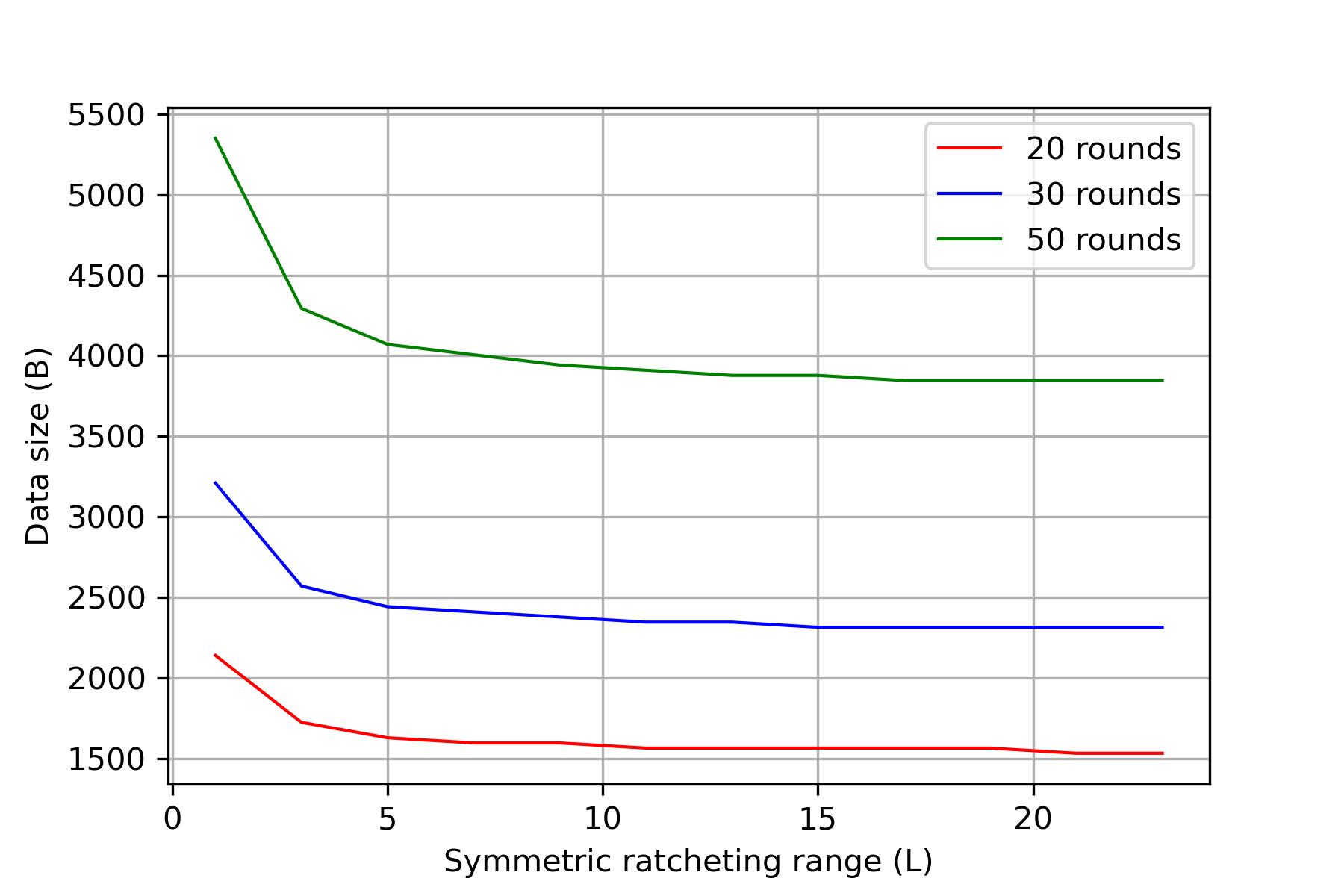}
        \caption{Server}
    \end{subfigure}
    \caption{On-chain data transmitted for round training in different symmetric ratchet ranges.}
    \label{fig: data_transmitted_on-chain}
\end{figure}

Moreover, Figure \ref{fig: data_transmitted_on-chain} illustrates the communication cost (in bytes) on the on-chain channel for both a single participant and the server across different rounds.
It is evident that the on-chain channel incurs significantly lower communication overhead compared to the off-chain channel, and this overhead decreases as the symmetric ratcheting range increases. 
This reduction occurs because fewer public key hash data are transmitted through the on-chain channel. 
Additionally, the server consistently transmits more data on-chain than the participant. 
This difference arises from the server’s involvement in both \textit{Feedback} and \textit{Publish} transactions, whereas participants are only responsible for an \textit{Update} transaction in each round.

%\begin{figure}[ht]
%    \centering
%\includegraphics[width=3.2in,height=2.3in]{Images/7.png}
%    \caption{Size of the data transmitted in a fixed 10 symmetric ratcheting range.}
%    \label{fig: size of data transmitted}
%\end{figure}

%\begin{figure}[ht]
%    \centering
%\includegraphics[width=3.2in,height=2.3in]{Images/8.png}
%    \caption{On-chain data transmitted for round training in different symmetric ratchet ranges.}
%    \label{fig: data transmitted on-chain}
%\end{figure}

\begin{comment}
\begin{figure}[ht]
    \centering
    \subfloat[Participant]{\includegraphics[width=1.6in,height=1.6in]{Images/client_onchain.png}}%
    \hspace{0.001\linewidth} % Add small horizontal spacing
    \subfloat[Server]{\includegraphics[width=1.6in,height=1.6in]{Images/server_onchain.png}}%
    \caption{On-chain data transmitted for round training in different symmetric ratchet ranges.}
    \label{fig: data_transmitted_on-chain}
\end{figure}
\end{comment}

%============================
\section{Conclusion}
\label{Conclusion and Future Work}
%============================
In this paper, we propose a post-quantum-based federated learning protocol with a hybrid architecture to provide both confidentiality and high performance in FL systems.  
We employed a combined approach that involved post-quantum and traditional cryptography.  
In addition, we utilized blockchain technology to address participant privacy and traceability concerns, considering the data included in transactions and the communication overhead of the on-chain channel.
The blockchain also facilitates key exchange authentication to establish sessions between the participants and the server in the PQBFL-based FL system. 
Our proposed protocol fulfills forward secrecy, post-compromise, and Harvest-Now, Decrypt-Later security as a distinct contribution to FL systems, in addition to the conventional security requirements of FL environments, such as protection against free-riding attacks.  
The performance of the PQBFL in terms of computational and communication costs demonstrated its feasibility for real FL systems.  
Finally, in future work, we aim to apply post-quantum homomorphic cryptography to mitigate FL data privacy risks, such as honest-but-curious servers and membership inference attacks.

%==================================================
\section*{Acknowledgement}
%\label{Acknowledgement}
%=================================================
This work is financed through national funds by FCT - Fundação para a Ciência e a Tecnologia, I.P., in the framework of the Project UIDB/00326/2020 and UIDP/00326/2020 and the Science and Technology Development Fund, Macau SAR. (File no. 0044/2022/A1).

\balance
\bibliographystyle{apacite}
%\bibliographystyle{apalike} % or you can use "apa" if available
\bibliography{ref}
\pagebreak
\begin{appendices}

\renewcommand{\thesection}{Appendix \Alph{section}}
\renewcommand{\thefigure}{A\arabic{figure}}
\setcounter{figure}{0}
\renewcommand{\thetable}{B\arabic{table}}
\setcounter{table}{0}
\definecolor{Alto}{rgb}{0.862,0.858,0.858}

\end{appendices}
\end{document}